\documentclass[11pt]{article}

\usepackage{newtxtext,newtxmath}
\usepackage{bm}

\usepackage{graphicx}

\usepackage[letterpaper,margin=1in]{geometry}

\renewenvironment{abstract} {\quotation} {\endquotation}

\date{}

\makeatletter
\renewcommand{\fnum@figure}{\textbf{Figure \thefigure}}
\renewcommand{\fnum@table}{\textbf{Table \thetable}}
\makeatother

\usepackage{scicite}

\usepackage{url}

\usepackage[T1]{fontenc}
\newcommand{\be}{\begin{equation}}
\newcommand{\ee}{\end{equation}}
\newcommand{\bea}{\begin{eqnarray}}
\newcommand{\eea}{\end{eqnarray}}

\newcommand{\vecv}{{\bm v}}

\newcommand{\veck}{{\bm k}}

\newcommand{\vecB}{{\bm B}}

\newcommand{\veckappa}{{\bm \kappa}}

\newcommand{\vecvarpi}{{\bm \varpi}}

\definecolor{red}{rgb}{0.8,0,0}
\definecolor{orange}{rgb}{0.8,0.2,0.0}
\definecolor{blue}{rgb}{0.3,0.0,0.8}
\definecolor{violet}{rgb}{0.4,0,0.4}
\definecolor{green}{rgb}{0,0.5,0.0}

\usepackage[normalem]{ulem}  

\def\scititle{Spin effects in superfluidity, neutron matter, and neutron stars
}
\title{\bfseries \boldmath \scititle}

\author{
	Armen Sedrakian$^{1,2\dagger}$,
	Peter B. Rau$^{3\dagger}$\and
	\small$^{1}$Frankfurt Institute for Advanced Studies,  60438
        Frankfurt am Main, Germany. \and
	\small$^{2}$Institute of Theoretical Physics, University of
         Wroc\l{}aw, 50-204  Wroc\l{}aw, Poland.\and
        \small$^{3}$Columbia Astrophysics Laboratory, Columbia University, New York, NY 10027, USA.\and
	\small$^\dagger$These authors contributed equally to this work.
}

\begin{document} 

\maketitle

\begin{abstract} \bfseries \boldmath
  We review selected aspects of the interior physics of compact stars,
  focusing on the microscopic and macroscopic manifestations of spin,
  magnetic fields, and nucleonic superfluidity and
  superconductivity. Spin statistics of fermions allow quantum
  degeneracy pressure to determine the stability and global properties
  of neutron stars, whose structure depends sensitively on the strong
  interactions among baryons in dense matter. Using a generic
  metamodeling framework based on an expansion of the nuclear energy
  density around the isospin-symmetric and saturation-density limits,
  we highlight how various lesser-known terms in this expansion affect
  compact-star observables and review multimessenger constraints on
  mass, radius, and moment of inertia. The influence of magnetic
  fields on dense matter is examined, showing that substantial effects
  in their structure require extremely strong fields, whereas lower
  fields are sufficient to affect their superfluid phases. At the
  mesoscopic scale, the coexistence of superfluid and superconducting
  components features vortex and flux-tube lattices, with pinning and
  mutual friction processes playing central roles in neutron-star
  rotational dynamics. We discuss unresolved issues concerning vortex
  structure, flux-tube configurations, and the origin of pulsar
  glitches and postglitch relaxation.  We also briefly address the
  possible emergence of deconfined quark phases in compact-star cores,
  including their color-superconducting properties, as well as the
  associated vortex structures and magnetic-field responses in such
  phases.
\end{abstract}

\noindent

\section {Introduction}
\label{sec:intro}

The year 2025 marks the centennial of the proposal of electron spin --
an intrinsic two-valued form of angular momentum -- by 
Uhlenbeck and Goudsmit~\cite{UhlenbeckGoudsmit1926}.  Their
work provided the first clear formulation of spin as an intrinsic
property of the electron, building on earlier experimental and
theoretical indications, including the anomalous Zeeman effect, the
Stern–Gerlach experiment, and early discussions of electron magnetic
moments by A.~Compton.  Originally introduced to explain the puzzling
doubling and splitting of atomic spectral lines, the concept of spin
soon revolutionized physics across all scales. From atomic structure
and chemistry to condensed matter, quantum field theory, and modern
quantum technologies, spin has become a cornerstone of our
understanding of nature. Although initially met with skepticism, the
concept soon found solid theoretical grounding. Pauli formalized its
quantum description through his matrices and the exclusion
principle~\cite{Pauli1925,Pauli1927}, while Dirac’s relativistic wave
equation~\cite{Dirac1928} revealed spin as an inherent and inevitable
consequence of relativistic quantum mechanics.

In astrophysics, the impact of spin is especially profound. Unlike
ordinary stars, neutron stars  are supported not by thermal pressure
but by the degeneracy pressure arising from the Pauli exclusion
principle and the spin-statistics connection for fermions. In this
sense, neutron stars represent nature’s most notable astrophysical
manifestation of fermionic spin: city-sized stellar remnants held up
against gravity primarily by the quantum property of spin in their
constituent neutrons (and, to a lesser degree, other baryons). What
began as an ad hoc solution to a spectroscopic puzzle thus became a
foundational pillar of modern physics, linking the microscopic
structure of elementary particles to the macroscopic stability of
stars.

Neutron stars are typically understood as objects composed primarily of nucleonic degrees of freedom, with an admixture of electrons (and muons at sufficiently high densities). They belong to the broader class of compact stars, all of which derive their stability from degeneracy pressure: (a) White dwarfs, characterized by lower average densities and substantially larger radii, are supported by the pressure of a cold, relativistic electron gas; (b) Hybrid and hyperonic stars, which can be regarded as subclasses of neutron stars, may be stabilized (in part) by the degeneracy pressure of cold, dense quark matter or hyperonic components, respectively, located deep within the stellar core and surrounded by an envelope of ordinary nuclear matter; (c) Strange stars are hypothesized to consist of nearly equal numbers of up, down, and strange quarks, with their stability arising from quark degeneracy pressure. These objects remain elusive, with no observational evidence to date supporting their existence. Note that white dwarfs are separated from neutron stars by an instability region in parameter space (e.g., central density versus stellar mass), whereas hybrid stars may or may not form distinct branch(es) relative to neutron (or hyperonic) stars. Strange stars are generally disconnected from other compact star families.

Thus, the centennial of electron spin is also a moment to reflect on and review its cosmic implications: the stability of compact stars, the fate of massive stellar evolution, and the extraordinary role played by a quantum mechanical principle, first invoked to explain the
fine details of atomic spectra, in producing the strongest gravity, most compact exotic stellar objects in the universe.

This review provides a concise and up-to-date account of the interior physics of compact stars, with particular emphasis on recent theoretical and observational developments, as well as on the fundamental role of spin in determining their structural and dynamical properties. Neutron star interiors may contain a significant—and in some cases even dominant—fraction of heavy baryons, such as hyperons and $\Delta$-resonances. These can be viewed as heavier counterparts of nucleons: hyperons carry strangeness, while $\Delta$-resonances correspond to higher-spin excitations. In what follows, we use the term ``neutron stars" to denote objects composed primarily of nucleonic degrees of freedom; however, we comment where appropriate on the modifications due to other degrees of freedom that become relevant.

Section~\ref{sec:EoS} outlines the equation of state (EoS)
of neutron-star matter and summarizes current astrophysical
constraints on neutron-star mass, radius, and moment of
inertia. Several representative numerical examples are presented to
illustrate generic trends. Section~\ref{sec:Bfields} examines the
microphysics of strongly magnetized neutron stars, highlighting the
influence of spin in such environments, especially the nontrivial
consequences of spin degeneracy lifting and Landau
quantization. Section~\ref{sec:Superfluidity} reviews superfluidity
and superconductivity in neutron-star interiors, emphasizing pairing
patterns that depend sensitively on the spin configuration of the
neutron and proton Cooper pairs in different density regimes. The
spin-1 triplet pairing and effects of the magnetic field on the
pairing patterns are discussed, together with the emergence of quantum
vorticity as a response to rotation and the magnetic field. Finally,
Section~\ref{sec:Astro} considers the macroscopic manifestations of
superfluidity in neutron stars and their astrophysical implications,
focusing on rotational irregularities such as glitches and long-term
precession phenomena. This review concludes with a discussion of pairing, spin, and superfluidity in quark matter in Sec.~\ref{sec:Quark_Superfluidity}, where we emphasize the analogies with, and differences from, nucleonic systems rather than presenting a fully independent treatment. Nevertheless, we provide key references to guide the interested reader toward more detailed studies of this topic.

The selection of topics and examples in this
review is, naturally, guided by the authors’ own research interests
and emphasizes the role of spin, which constitutes the central theme
of this work. Furthermore, the bibliography is intended to be
illustrative rather than exhaustive, highlighting representative
contributions rather than offering a comprehensive survey of the
field.

\section{Spin and the EoS of neutron stars}
\label{sec:EoS}

\subsection{Spin, quantum statistics and degeneracy pressure}
\label{sec:EoS:A}

At the most fundamental level, the spin–statistics connection ensures the existence of compact stars through degeneracy pressure. Neutrons, as spin-$1/2$ fermions, obey Fermi–Dirac statistics, and their degeneracy pressure counteracts gravitational collapse. Additional baryonic species, such as protons, hyperons, and baryon resonances, also contribute—though typically at a smaller level, depending on the star’s composition. While hyperons, like nucleons, carry spin $1/2$, $\Delta$-resonances have spin $3/2$; both are fermions and contribute to the total degeneracy pressure, with their impact entering through different degeneracy factors in the phase-space 
integrals as well as from the specifics of their distinct dynamical equations (Dirac equation or Rarita--Schwinger equation) and strong and electroweak interactions with their environment~\cite {Lattimer2004Sci,Sedrakian2023PrPNP}. 

Although the model of a non-interacting degenerate Fermi gas is unrealistic, it is worthwhile
to briefly look at it, firstly, because it was the first model to
describe degenerate compact stars, and secondly, it exhibits the
importance of the spin and Fermi statistics in creating the degeneracy
pressure~\cite{Tolman1939,Oppenheimer1939}.  At zero temperature,
the picture is such that all the states below the Fermi energy
$E_{F,n}$, or the associated momentum $p_{F,n}$ are occupied, whereas
all the states above are free. The relativistic spectrum of neutrons
$E_{F,n}= \sqrt{m_n^2 c^4+ p_{F,n}^2c^2}$, where $m_n$ is the (bare or
effective) neutron mass, $p_{F,n}$ is the Fermi momentum, is useful to
express in terms of the dimensionless relativity parameter for
neutrons $x_n$ via
$E_{F,n}=m_n c^2\cosh \left({x_n}/{4}\right),$
$x_n  = 4\sinh^{-1} \left({p_{F,n}}/{m_nc}\right).$
Thus, $x_n$ encodes the degree of relativism of the neutrons:
small $x_n$ corresponds to the non-relativistic limit, while large
$x_n$ corresponds to the ultrarelativistic regime. Computing the energy
density by integrating the distribution function of neutrons over momentum
space up to the Fermi momentum, and similarly, the pressure using the
standard thermodynamics formulas, one finds at zero temperature~\cite{Landau1980}
\bea\label{eq:rho_n}
{\cal E}_n &=&  K_n \left(\sinh x_n - x_n \right),\\
\label{eq:P_n}
P_n &=& K_n \left( \frac{1}{3}\sinh x_n - \frac{8}{3} \sinh\left(\frac{x_n}{2}\right)
  + x_n \right),
\eea
where $ K_n = {m_n^4 c^5}/{32 \pi^2 \hbar^3},$ sets the overall
density and pressure scale. It has dimensions of energy density and
arises naturally from the phase-space volume of fermions (counting
states per unit volume in momentum space with a factor 2 for spin
degeneracy assuming spin 1/2 fermions).  Together,
Eqs.~\eqref{eq:rho_n} and \eqref{eq:P_n} describe the EoS of cold,
degenerate neutrons. The limiting cases of non-relativistic $x_n\ll 1$
and ultra-relativistic $x_n \gg 1$ gas are easily recovered by
expanding these equations respectively in $x_n$ and $x_n^{-1}$,
showing the polytropic forms of the EoS in these two limits
\be
P_n \propto
    \begin{cases}
    {\cal E}_n^{5/3}, & \quad x_n\ll 1,
    \\
    {\cal E}_n^{4/3}, &  \quad x_n\gg 1.
    \end{cases}
\ee
Although real neutron-star matter involves strong nuclear
interactions, the relativistic Fermi gas model highlights the role of
the Pauli principle and spin–statistics in supporting neutron stars
against gravity~\cite{Oppenheimer1939}. Corrections due to interactions are, however,
significant. The predictions for the global parameters of the neutron
stars by the degenerate ideal gas EoS are unrealistic, in particular,
the maximum mass of a sequence of compact stars turns out to be too
low and in contradiction to the observations.

\subsection{Astrophysical limits}

Pulsar timing provides precise neutron star mass measurements through
Keplerian and relativistic parameters of binary systems, including
neutron star–neutron star and neutron star–white dwarf pairs. Although
neutron star–black hole binaries have been detected in gravitational
waves, they have not yet yielded strong constraints on the neutron
star structure. The Shapiro delay method—based on the extra time radio
pulses take to traverse the gravitational field of a companion—has
been key in measuring the masses of massive pulsars in a binary with a
white dwarf~\cite{Shapiro1964}.  The first such measurement, for PSR
J1614-2230, yielded a mass of
$1.908(16)\,M_\odot$~\cite{Demorest:2010bx,Arzoumanian2018,Cromartie:2019kug}. The
second, PSR J0348+0432, was found to have $M = 2.01 \pm 0.04\,M_\odot$
using combined timing and optical
modeling~\cite{Antoniadis:2013pzd}. The most massive to date known
neutron star, PSR J0740+6620, has a measured mass of
$2.08 \pm 0.07\,M_\odot$ \cite{Fonseca:2021wxt}.

In general relativity, stability requires that stellar mass increase
with central density according to the Bardeen--Thorne--Meltzer
criterion~\cite{Bardeen1966}, which is equivalent to the requirement
that the oscillation modes of the star remain stable on the ascending
branch of mass-central density curve. Measurements of massive
pulsars, thus firmly establish that the maximum mass of a neutron
star should not be lower than the observed one (currently
$M\simeq 2\,M_\odot$) setting a lower bound on the maximum mass
allowed by any viable EoS.

Gravitational-wave observations from binary neutron star mergers now provide complementary constraints. The LIGO--Virgo
detections of GW170817 and GW190425~\cite{AbbottPhysRevX.9.011001,Abbott2020} enabled the
measurement of tidal deformability, which relates the induced
quadrupole moment $Q_{ij}$ to the external tidal field $E_{ij}$ via
\bea
Q_{ij} = -\lambda E_{ij} = -  \frac{2}{3}k_2 R^5 E_{ij},
\eea
where $\lambda$ is the tidal deformability parameter, $R$ is the
stellar radius, and $k_2$ the Love
number~\cite{flanagan2008constraining,Hinderer:2007}. It is seen that
$\lambda \propto R^5$ showing its sensitivity to the EoS. Analyses
under low-spin priors yielded effective (dimensionless) deformabilities
$\tilde{\Lambda} \simeq 300^{+500}_{-190}$ for GW170817 and an upper
limit $\tilde{\Lambda} \le 600$ for GW190425, implying radii
$R \simeq 11-13.5$~km for typical $1.4\,M_\odot$
stars~\cite{AbbottPhysRevX.9.011001,Abbott2020}.

X-ray pulse profile modeling with the orbital NICER instrument has
provided independent mass–radius constraints. For PSR J0030+0451,
analyses gave $M = 1.34$–$1.44\,M_\odot$ and
$R = 12.7$–$13.0$~km~\cite{NICER:2019a,NICER:2019b}, while for PSR
J0740+6620 NICER found
$R = 12.4$–$13.7~\mathrm{km}$~\cite{NICER:2021a,NICER:2021b} and a
mass consistent with that obtained from radio-timing observations
quoted above. Importantly, these constraints—like
those inferred from gravitational-wave observations—do not correspond
to single data points with $1\sigma$ error bars, but rather to
posterior probability regions in a two-dimensional parameter space
(e.g., the mass–radius or tidal deformability planes). Recent
reanalyses (combining NICER with XMM-Newton data from 2023–2024)
suggest radii that are consistent with or slightly smaller than
earlier estimates~\cite{Vinciguerra2024}, with values around
$R \approx 12.3 \pm 0.6$~km, thereby favoring an EoS featuring a
relatively steep rise in pressure beyond nuclear density to support
massive neutron stars. In summary, combining pulsar timing,
gravitational-wave, and NICER measurements jointly constrains the EoS:
the maximum mass must exceed $2M_\odot$, while the radii of
$\sim 1.4M_\odot$ to $\sim 2.0M_\odot$ stars cluster around
$\sim 12$–$13$~km, implying a weak dependence of the radius on the
stellar mass.

Binary neutron star systems containing two pulsars provide an
exceptional natural laboratory for testing the predictions of
relativistic astrophysics~\cite{Kramer2021}. These rare systems enable a variety of
precision tests of general relativity, including measurements of
orbital decay due to gravitational wave emission, Shapiro delay, and
relativistic periastron precession. In addition to their role in
testing gravity, such systems also offer valuable constraints on the
EoS of dense nuclear matter, since pulsar timing allows for highly
accurate determinations of stellar masses and, potentially, moments of
inertia.

A particularly notable example is the double pulsar system PSR
J0737$-$3039, which consists of two active, radio-emitting neutron
stars. This system is unique because both stars are detectable as
pulsars, allowing an unprecedented level of orbital and timing
precision. Over long observational baselines, relativistic
effects--especially the periastron advance--can be used to extract the
moment of inertia of pulsar A, denoted $I_A$, solely from measurable
post-Keplerian parameters, without relying on theoretical modeling of
the EoS. This possibility was first highlighted by Lattimer and
Schutz~\cite{LattimerSchutz2004}, who demonstrated that a measurement
of $I_A$ to $10\%$ accuracy could directly translate into strong
constraints on the neutron-star EoS.  Subsequent statistical analyses
by Kramer et al.~\cite{Kramer2021} have derived posterior probability
distributions for the component masses and the moment of inertia of
pulsar A, yielding the upper bound
$
I_A < 3.0 \times 10^{45}\, \mathrm{g\,cm}^2
$
at the 90\% confidence level.  Continued timing observations of PSR
J0737$-$3039 over the coming decade or so are expected to refine the
measurement of relativistic parameters to the point where the
empirical uncertainty on $I_A$ becomes comparable to that obtained
from X-ray pulse-profile modeling by the NICER mission. Such a
measurement would represent a major step forward, providing a purely
dynamical, model-independent determination of a neutron star’s moment
of inertia--and, by extension, a stringent new test of dense-matter
physics.

\subsection{Equation of state}
\label{sec:equation of state:B}

The role of spin in unpolarized neutron star matter, which is a good
approximation for the magnetic fields typical of ordinary pulsars,
$B \sim 10^{12}-10^{13}$~G, is straightforward: it effectively doubles
the phase space available to fermions at a given energy state,
corresponding to the spin-up and spin-down configurations. From a
microscopic perspective, however, the interaction channels described
by the partial wave expansion of the nuclear potential obey
spin-isospin selection rules, so spin plays a nontrivial role. 

In
low-density neutron matter, spin-singlet $^1S_0$ interactions between neutrons dominate, as well as similar interactions between protons
which have much lower densities. In contrast, in symmetric nuclear
matter, the dominant low-density interaction is the coupled-channel spin-triplet
$^3S_1$-$^3D_1$ interaction, which is, however, suppressed at the
extremely large isospin asymmetries present in neutron stars and
enforced by  weak interactions which establish \textit{beta-equilibrium}. At higher densities, $P$-wave
channels become important; these contain both spin-singlet and
spin-triplet contributions, specifically the spin-singlet $^1P_1$ and
spin-triplet $^3P_0$, $^3P_1$, and $^3P_2$-$^3F_2$ channels, the last 
of which dominates the pairing interaction
between neutrons at densities above the saturation density $\rho_{\text{sat}}$.

To gain a qualitative understanding of these interactions, it is
useful to express the energy per baryon of nuclear matter as an expansion in a
Taylor series
\bea
\label{eq:Taylor_expansion}
E(\chi, \delta) &\simeq& E_{\rm sat} + \frac{1}{2} K_{\rm sat}\,
\chi^2 + \frac{1}{6} Q_{\rm sat}\, \chi^3 + E_{\rm sym}(\chi) \delta^2
+ \dots\nonumber\\
\eea
with
\bea
\label{eq:Esym_expansion}
E_{\rm sym} &=& J_{\rm sym} + L_{\rm sym} \chi +
\frac{1}{2} K_{\rm sym} \chi^2 + \frac{1}{6} Q_{\rm sym}\chi^3 + \cdots\nonumber\\
\eea
where $\chi=(\rho-\rho_{\text{sat}})/3\rho_{\text{sat}}$ and $\delta = (\rho_{n}-\rho_{p})/\rho$, with $\rho_n$ and $\rho_p$ being the neutron and proton densities.  We note that Eq.~\eqref{eq:Taylor_expansion}, apart from straightforward expansion in density,  also corresponds to an expansion in the isospin asymmetry parameter $\delta$, truncated at quadratic order (the parabolic approximation). {While this approximation is well justified near symmetric nuclear matter, its validity becomes less controlled for highly neutron-rich matter $(\delta \sim 1)$, where higher-order terms in $\delta$ may contribute non-negligibly. Its use in the neutron-star context is nevertheless supported by empirical evidence from microscopic models, which indicate that the quadratic approximation remains reasonably accurate even at large isospin asymmetry~\cite{Gandolfi2012}.
The symmetry energy expansion can incorporate hyperons by including
the strangeness quantum number as an expansion parameter and
redefining the isospin asymmetry parameter~\cite{Yang:2025}.

The coefficients of the expansion of the symmetric nuclear matter   are  the {\it binding
  energy} $E_{\rm sat}$, {\it incompressibility} $K_{\text{sat}}$, and the {\it skewness} $Q_{\text{sat}}$;  the coefficients of the density expansion of 
 $E_{\rm sym}(\chi)$ are the {\it symmetry energy} $J_{\text{sym}}$, the {\it slope and curvature parameters} $L_{\text{sym}}$
 and $K_{\text{sym}}$, and the {\it skewness of symmetry energy}  $Q_{\rm sym}$.
   These  are defined explicitly as 
\bea\label{eq:isoscalar_coeff}
K_{\text{sat}} =9 \rho_{\text{sat}}^2 \frac{d^2 E(\chi)}{d  \rho^2},\quad 
Q_{\mathrm{sat}}=27 \rho_{\text{sat}}^3 \frac{d^3 E(\chi)}{d \rho^3},
\eea
and
\bea\label{eq:isovector_coeff}
L_{\mathrm{sym}}  &=&3 \rho_{\text{sat}} \frac{d E_{\mathrm{sym}}(\chi) }{d \rho},\quad
K_{\mathrm{sym}} =9 \rho_{\text{sat}}^2 \frac{d^2 E_{\mathrm{sym}} (\chi)}{d \rho^2},
\eea
where all derivatives are evaluated at $\rho=\rho_{\rm sat}$.  The
expansion \eqref{eq:Taylor_expansion} is rooted in the infinite-matter
limit of the Bethe--Weizs\"acker nuclear mass formula. As such, its
low-order coefficients are well constrained by experimental data on
finite nuclei. In particular, the binding energy per nucleon at
saturation, $E_{\rm sat} \simeq -16.0 \pm 1$~MeV, the nuclear symmetry
energy at saturation, $J_{\rm sym} \simeq 32.0 \pm 2$~MeV, and the
saturation density itself,
$\rho_{\rm sat} \simeq 0.15-0.16\ \mathrm{fm}^{-3}$, are among the
most reliably determined quantities.  Higher-order coefficients in the
density expansion, both in the isoscalar \eqref{eq:isoscalar_coeff}
and isovector \eqref{eq:isovector_coeff} channels are less constrained
by direct experiments. Instead, they have been studied using a
combination of statistical Bayesian
analyses~\cite{Imam2022,Xie:2020kta,Li:2025oxi,Li:2025vhk},
metamodeling approaches~\cite{Margueron:2018eob}, and multimessenger
observations of compact stars~\cite{Marczenko:2020jma,Koehn2025}.  In
general, the validity of the Taylor expansion
\eqref{eq:Taylor_expansion} is limited to the condition $\chi < 1$,
which corresponds to densities to a few times $\rho_{\rm sat}$
depending on the parametrization of density functional for nucleonic
models, but this estimate may change if additional degrees of freedom,
such as hyperons or $\Delta$-resonances, become relevant, or if
deconfinement phase transition to quark matter occurs.

Despite considerable progress, $K_{\rm sat}$ is still estimated within
a relatively broad range,
$200 \lesssim K_{\rm sat} \lesssim 300$~MeV~. Heavy-ion collision
experiments favor a relatively soft EoS, implying lower
$K_{\rm sat}$, although such conclusions remain model-dependent.
The skewness parameter $Q_{\rm sat}$, which characterizes the third
derivative of the energy with respect to density, remains largely
unconstrained. Estimates in the literature vary widely, typically
within $-600 \lesssim Q_{\rm sat} \lesssim 1000$~MeV. It is important
to emphasize once again that the above constraints pertain specifically to
nucleonic matter. Their applicability to dense matter containing
heavier baryons, such as hyperons or $\Delta$ resonances, is uncertain and subject to ongoing investigation~\cite{Sedrakian2023PrPNP}.

\subsection{Mass, Radius, and Moment of Inertia}

The static properties of compact stars, in the simplest case assuming
spherical symmetry (i.e., negligible rotation, and weak magnetic
fields) are obtained by integrating the Tolman-Oppenheimer-Volkoff
 equations~\cite{Tolman1939,Oppenheimer1939}. These equations
represent the solution of Einstein’s field equations for a spherically
symmetric mass distribution in hydrostatic equilibrium. To facilitate
comparison between theoretical predictions and observations, it is
often useful to present results in terms of directly observable
quantities, such as the stellar mass, radius, moment of inertia, and
spin frequency.

Figure~\ref{fig:equation of state_QL} shows representative EoS of
neutron star matter with nucleonic degrees of freedom only
constructed from a covariant density functional where the parameters
$Q_{\rm sat}$ and $L_{\rm sym}$ are varied to illustrate their effect
on the pressure of matter~\cite{Li:2023bid}. The corresponding
solutions of the Tolman-Oppenheimer-Volkoff equations are shown in
Fig.~\ref{fig:TOV_QL} (a), \{where the mass-radius relations for
neutron star matter with nucleonic degrees of freedom only are
displayed along with the current astrophysical constraints.
\begin{figure}[t]
\centering
\includegraphics[width = 0.7\textwidth]{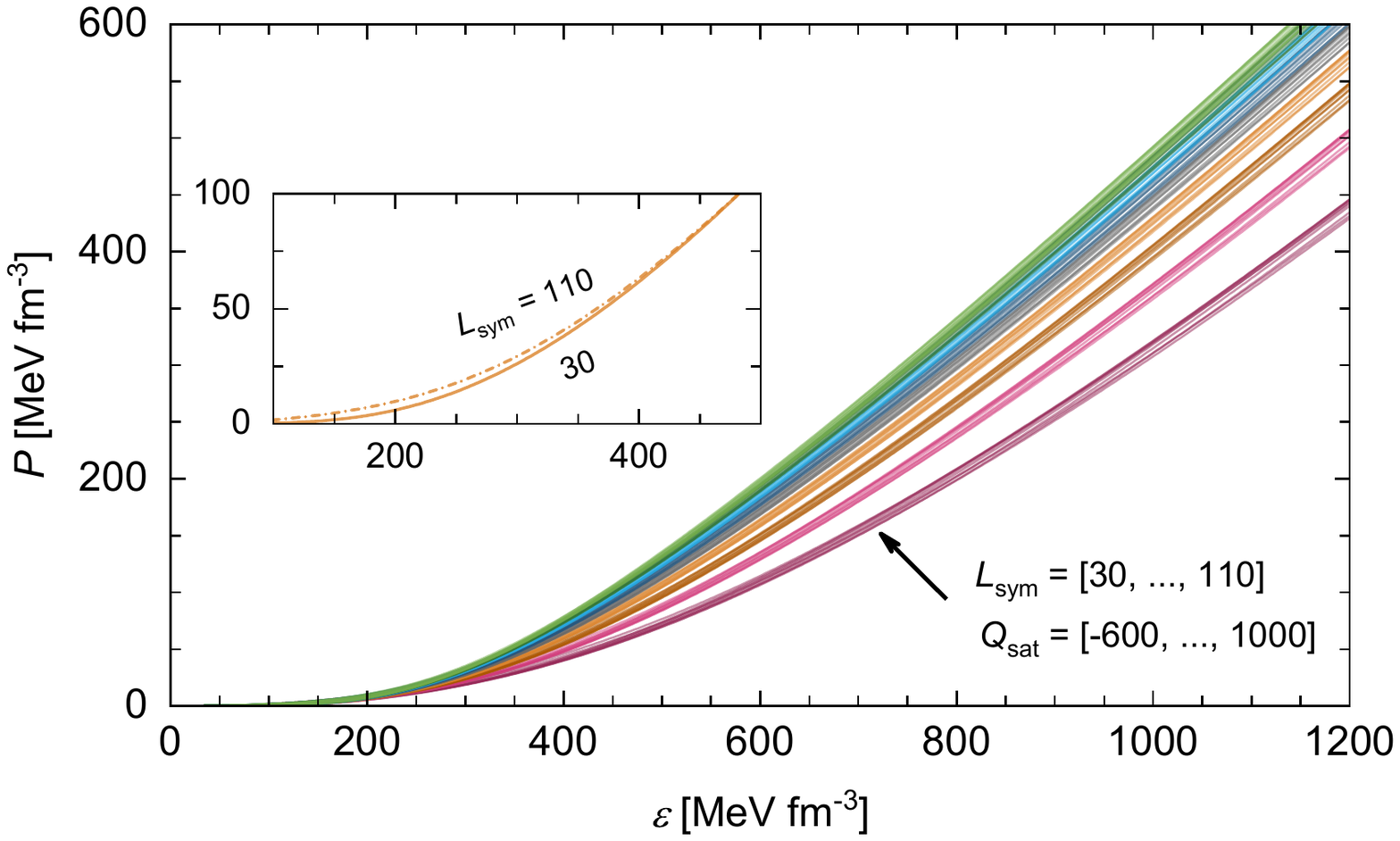}
\caption{ {\bf Variation of the nuclear equation of state with
    $Q_{\rm sat}$ and $L_{\rm sym} $ in covariant density functional
    models.}  Family of EoS generated using covariant density
  functional models with the parameters varied over the ranges
  $Q_{\rm sat} \in [-600,1000]$~MeV and
  $L_{\rm sym} \in [30,\,110]$~MeV (distinguished by color); other
  parameters of covariant density functional are fixed to that of
  DDME2 parametrization; for details see Ref.~\cite{Li:2023bid}. The
  inset highlights the impact of $L_{\rm sym}$ on the low-density
  region of the EoS for illustration.}
\label{fig:equation of state_QL}
\end{figure}
\begin{figure}[thb]
\centering
\includegraphics[width = 16cm]{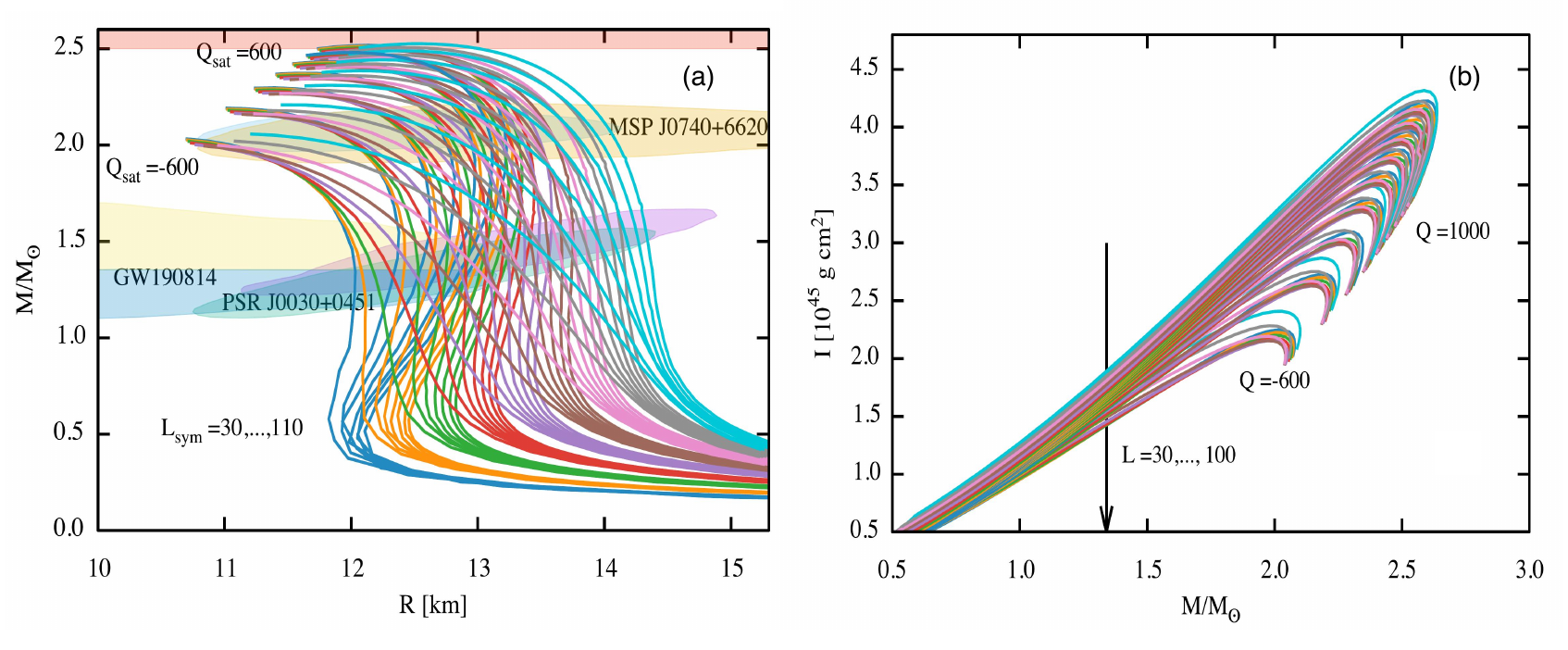}
\caption{
{\bf Mass–radius and moment-of-inertia–mass relations for nucleonic equations of state compared with multimessenger astrophysical constraints.}
  ({\bf a}) Mass–radius for nucleonic EoS models corresponding to
  different $Q_{\rm sat}$ and $L_{\rm sym}$ pairs (in MeV)~\cite{Li:2023bid}. The
  colored regions indicate the 90\% confidence interval (CI) ellipses
  from the two NICER analyses for PSR J0030+0451 and PSR J0740+6620
  \cite{NICER:2019a,NICER:2019b, NICER:2021a,NICER:2021b}, the 90\% CI
  regions for the two compact stars involved in the gravitational-wave
  event GW170817~\cite{AbbottPhysRevX.9.011001}  and the 90\%CI for the mass of
  the secondary component in GW190814~\cite{LIGO-Virgo2020}.
  ({\bf b}) Moment of inertia vs mass for the same EoSs as in (a)
  for the indicated range of $Q_{\rm sat}$ and
  $L_{\rm sym}$ (in MeV); the upper limit $I_A\le 3\times 10^{45}$ g
  cm$^2$ at 90\% CI on the moment of inertia extracted from PSR
  J0737$-$3039 system (shown by vertical arrow) is well satisfied for
  the collection of EoSs, given that the pulsar mass
  $m_A \simeq 1.34\,M_{\odot}$.  
  }
\label{fig:TOV_QL}
\end{figure}
By systematically varying $Q_{\rm sat}$ and $L_{\rm sym}$, one can
directly relate microphysical properties of nuclear matter to
macroscopic observables, including the stiffness of the EoS, neutron
star radii, tidal deformabilities, and gravitational-wave signatures
from binary mergers for a given composition. Of course, such a
procedure requires additional physical input—namely, assumptions about
the composition of the deep interior of the star, whether purely
nucleonic, admixed with heavy baryons, or involving phases featuring
quark matter.

This approach provides a powerful pathway to constrain nuclear physics
models through astrophysical data, establishing a direct connection
between dense matter microphysics and observable neutron star
phenomena. For completeness, we show in Fig.~\ref{fig:TOV_QL} (b), the
moment of inertia as a function of mass for the same collection of
EoS. The upper limit on the moment of inertia of the A pulsar
$I_A\le 3\times 10^{45}$ g cm$^2$ of PSR J0737$-$3039 is fully
consistent with the entire collection of the EoS given that its mass
was estimated to be $m_A \simeq 1.34\,M_{\odot}$.

The possible appearance of heavy baryonic degrees of freedom—such as
hyperons and $\Delta$-resonances—in dense matter has important
implications for the global properties of neutron stars. Their
inclusion is naturally accommodated within covariant density
functional approaches, which provide a flexible framework constrained
by both laboratory data and astrophysical observations. The onset of
hyperons at supranuclear densities generally softens the equation of
state, leading to a reduction in the maximum mass of neutron stars—a
tension commonly referred to as the {\it hyperon puzzle}, given the
observation of compact stars with masses above
$2\,M_\odot$~\cite{Bombaci2017,Oertel2017,Sedrakian2023PrPNP}.

Modern covariant density functional models address this issue by
tuning the underlying couplings (e.g., within $SU(6)$ or $SU(3)$
symmetry schemes), thereby enabling sufficiently stiff equations of
state that support massive stars while still allowing hyperonic
degrees of freedom. In typical scenarios consistent with $SU(6)$
symmetry, hyperons can constitute up to $\sim 20\%$ of the stellar
core composition, with $\Lambda$ hyperons becoming dominant at high
densities. In contrast, models adjusted to reproduce very massive
objects tend to suppress the hyperon fraction to only a few percent,
effectively yielding nucleonic stars.

Beyond static structure, heavy baryons also influence rotational
properties, magnetic configurations, and thermal evolution. In
particular, hyperons enable fast cooling via direct Urca
processes~\cite{Prakash1992}, although this can be mitigated by
hyperonic pairing, leading to a characteristic mass-dependent cooling
behavior~\cite{Fortin2021,Raduta2019}. Finite-temperature effects,
relevant for supernovae and binary neutron star mergers, further
promote the appearance of hyperons at lower densities. Overall, the
inclusion of heavy baryons enriches the phenomenology of neutron
stars, but their quantitative impact remains sensitive to poorly
constrained interactions and many-body effects, making them a key
target for future observational and theoretical
studies~\cite{Providencia2019FrASS}.

Beyond purely nucleonic and hypernuclear stars, compact stars may also
contain deconfined quark matter in their inner cores, leading to
so-called hybrid stars. This possibility is particularly interesting
because first-order phase transitions in dense matter can leave direct
imprints on global stellar properties. If the transition from hadronic
to quark matter involves a sufficiently large jump in energy density,
the resulting softening and subsequent re-stiffening of the equation
of state can generate new stable branches in the mass–radius
diagram. In the simplest case, this gives rise to ``twin''
configurations, where two stars of the same mass have different radii
and internal compositions, one being purely hadronic and the other
containing a quark core~\cite{Alford2013,Christian2018,Li2025JCAP}. If
the phase structure of dense QCD matter is more intricate, with
sequential transitions between distinct quark phases, the mass–radius
relation may exhibit an even richer pattern, including additional
stable branches and higher-order multiplets~\cite{Alford2017}. More
generally, even in model-agnostic approaches, the presence of
deconfined quark matter can manifest itself through nontrivial
features in the EoS, such as a rapid variation of the speed of sound,
which are consistent with current astrophysical constraints and can
significantly impact masses, radii, and tidal
deformabilities~\cite{Tan2020}. In this sense, hybrid stars provide a
natural extension of the discussion of heavy-baryon degrees of
freedom: while hyperons soften the equation of state within the
hadronic description, deconfinement introduces qualitatively new
phases whose onset and stiffness can reshape the global properties of
compact stars in a potentially observable way.

We defer a detailed discussion of pairing, spin, and superfluidity in
quark matter to Sec.~\ref{sec:Quark_Superfluidity}, after presenting
the physics of nucleonic phases. This ordering allows us to use the
framework developed for nucleonic superfluids as a basis for
comparison, highlighting both the analogies and the key differences
that arise in quark matter, and providing a coherent entry point to
the more specialized literature on this topic.

\section{Spin and magnetic field}
\label{sec:Bfields}

For the class of strong-magnetic-field compact stars, known as magnetars, the degeneracy with respect to the spin  is lifted by the magnetic field--spin interaction~\cite{Turolla2015,Adhikari2026PrPNP}. Magnetars appear as
anomalous X-ray pulsars, soft gamma ray repeaters, and as at least some of the fast
radio bursts, which can be endowed with exceptionally strong surface
magnetic fields, typically in the range
$10^{14}$–$10^{15}$~G. Theoretical models further indicate that the
equilibrium configurations of compact stars can support even stronger
internal magnetic fields, up to
$B \lesssim 10^{18}$–$10^{19}$~G~\cite{Bocquet1995, Broderick2000}.  In this range of
fields, observable effects on global parameters such as the stellar
mass, radius, and moment of inertia occur. This is also a regime,
where the star is close to the limit of gravitational stability
predicted by virial-theorem arguments and confirmed by
general-relativistic stellar models~\cite{Bocquet1995}.

Microscopically, the influence of magnetic fields on the EoS of dense
matter becomes significant when the magnetic field energy associated
with charged or magnetized particles approaches the characteristic
Fermi energies of the system.  Because the Fermi energy increases from
the low-density outer layers toward the dense core, stronger magnetic
fields are required to significantly modify the EoS at greater depths
within the star~\cite{Cardall2001,
  Thapa:2021kfo,Peterson:2023bmr,Most2025ApJ}.

In a charge-neutral, beta-equilibrated neutron-star core, the charged
constituents—electrons, muons, and protons—are the primary carriers
affected by the field. In a magnetic field, the transverse motion of
charged particles is quantized into Landau
levels~\cite{Lai2001,Harding2006}. For relativistic electrons,
assuming a $g$-factor of exactly 2 and a magnetic field in the
$z$-direction, the energy spectrum takes the form
\be E_{n}(p_z) =
\sqrt{ p_z^2 c^2 + m_e^2c^4\left(1 + 2 n \frac{B}{B_{c,e}}\right) },
\label{eq:E_Landau_e}
\ee
where $n = 0,1,2,\dots$ labels the Landau level, and $p_z$ is the momentum along the magnetic field
and $B_{c,e}$ is a critical field defined below.  Correspondingly, at zero temperature, the electron number density becomes
 \bea
 n_e=\frac{eB}{2\pi^2\hbar c}\sum_{n=0}^{n_{\rm max}} g_n p_{F,n},
 \eea
 where $g_n=2-\delta_{n0}$ is the spin degeneracy factor, $p_{F,n}$ denotes the Fermi momentum along
 the magnetic-field direction for the $n$-th Landau level,  and $n_{\rm max} = \lfloor p_{F,0}^2B_{ce}/(2m_e^2c^2B)\rfloor$ is the highest occupied Landau level.

Relativistic effects associated with Landau quantization become important once the cyclotron energy, $e\hbar B / (m c)$, becomes comparable to the particle’s rest energy $m c^2$. This defines a critical field strength for each particle species. For electrons, the critical field is
\bea
B_{c,e} = \frac{\hbar}{e\lambda_e^2} = 4.414\times10^{13}~\mathrm{G},
\eea
where $\lambda_e = \hbar / (m_e c) \simeq 400~\mathrm{fm}$ is the Compton wavelength of the electron. For protons, the critical field is much higher, $B_{c,p}\sim 10^{20}~\mathrm{G}$, and likely exceeds the hydrostatic stability limit.

Quite generally, Landau quantization reduces the phase space available for charged particles by confining them to a limited number of occupied Landau levels, especially when $B \gtrsim B_{c,e}$ and only a few levels are populated. This modification of the density of states alters the pressure and energy density, typically leading to a softening of the EoS. In addition, it affects the composition of matter through changes in the charge fraction $Y_Q(n_B) = n_Q/n_B$, where $n_B$ is the baryon density and $n_Q$ is the net charge density carried by all charged constituents, such as protons, electrons, muons, or, in the case of deconfined matter, quarks. This quantity characterizes how the electric charge content evolves with density under the constraints of charge neutrality and beta equilibrium.

The underlying reason is that the modified density of states for electrons (and, at sufficiently high fields, other charged species) alters the conditions of charge neutrality and beta equilibrium. As a result, the equilibrium proton and lepton fractions can deviate from their field-free values, particularly at low densities where Landau level quantization is most pronounced. These changes can, in turn, influence macroscopic stellar properties, including the maximum mass and radius, and are therefore particularly relevant for modeling strongly magnetized neutron stars (magnetars); for a broad review of these aspects see Ref.~\cite{Adhikari2026PrPNP}.

In the same field-strength range, additional effects arise from the anomalous magnetic moments of nucleons. Even though neutrons are electrically neutral, their intrinsic magnetic dipole moment couples directly to the magnetic field and contributes to both the energy density and the composition of matter. The corresponding energy shift associated with this coupling is of order $|\kappa_p + \kappa_n|\, B$,
which modifies the baryon single-particle energies and, consequently, the conditions for beta equilibrium in sufficiently strong magnetic fields. The experimentally determined anomalous magnetic moments of protons and neutrons are
\bea
\kappa_p = \mu_N \left( \frac{g_p}{2} - 1 \right), \qquad
\kappa_n = \mu_N \frac{g_n}{2},
\eea
where $\mu_N = e\hbar / (2m_p)$ is the nuclear magneton, and $g_p = 5.58$ and $g_n = -3.82$ are the Land\'e $g$-factors for the proton and neutron, respectively.

More explicitly, the single-particle spectrum for the baryons acquires a spin-dependent splitting in the presence of a magnetic field. Since the protons are Landau quantized similarly to the electrons, their spectrum differs from that of the neutrons in this way in addition to $\kappa_p\neq\kappa_n$: these spectra are, ignoring nuclear interactions~\cite{Broderick2000}
\begin{align}
E_{p,s}(p_z) ={}& \sqrt{p_z^2 c^2 + \left[ M_{p,n,s}(B) c^2 + s\kappa_pB \right]^2 },
\label{eq:E_Landau_p}
\\
M_{p,n,s}(B) ={}& m_p\sqrt{ 1+2\left(n+\frac{1}{2}+\frac{s}{2}\right)\frac{B}{B_{c,p}} },
\\
E_{n,s}(p_z,p_{\perp}) ={}& \sqrt{p_z^2 c^2 + \left[\sqrt{m_n^2c^4+p_{\perp}^2c^2}+s\kappa_nB \right]^2 },
\label{eq:E_Landau_n}\nonumber\\
\end{align}
for $s=\pm 1$, $n=0,1,2,...$ in the definition of $M_{p,n,s}(B)$, and where $p_{\perp}$ is the momentum in the plane perpendicular to the magnetic field. The spin-dependent term lifts the degeneracy between spin-up and spin-down states and leads to a partial spin polarization of the baryonic component. This effect becomes appreciable when the magnetic energy scale $\kappa_b B$, $b=p,n$, is comparable to the Fermi energy, which typically requires fields $B \gtrsim 10^{17}-10^{18}\,\mathrm{G}$. As a result, both the neutron and proton populations can become spin-polarized, altering the energy density and pressure, as well as shifting the beta-equilibrium condition through modified chemical potentials.

For comparison, the effect of the magnetic field on electrons is qualitatively different and typically much stronger at lower field strengths due to Landau quantization. As discussed above, the electron single-particle spectrum becomes discretized in the plane perpendicular to the field,
see Eq.~\eqref{eq:E_Landau_e},
with a corresponding reduction of the available phase space when only a few Landau levels are occupied. This leads to oscillatory behavior in thermodynamic quantities as a function of density or magnetic field strength, known as de Haas--van Alphen oscillations, and significantly modifies the electron chemical potential.

Together, these effects--Landau quantization for charged particles and spin polarization from anomalous magnetic moments for baryons--alter the composition of matter, including the charge fraction $Y_Q(n_B)$, by modifying the beta-equilibrium condition and the relative populations of particle species. While Landau quantization dominates at relatively lower field strengths $B \gtrsim 10^{13}-10^{14}\,\mathrm{G}$, the anomalous magnetic moment effects become important only at much higher fields, characteristic of the most strongly magnetized neutron stars.

For typical neutron star core densities, nucleon Fermi energies lie in
the range of a few to several tens of MeV above the rest mass energy. Therefore, once
$B/B_{c,e}\gtrsim 10^5$, the anomalous magnetic moment contribution becomes comparable to
these energies, and their impact on matter composition and pressure
cannot be neglected. In this extreme regime, the magnetic coupling of
nucleon spins leads to complete spin polarization of neutrons. The alignment of neutron spins enhances the pressure contribution from the neutron sector, producing an overall stiffening of the EoS that counteracts, and eventually dominates, the softening induced by Landau quantization. 

Thus, at the highest magnetic field strengths relevant for compact
stars, the net effect on the EoS results from a competition between
two mechanisms: (a) Landau quantization, which reduces the pressure by
restricting charged particles to lower Landau levels, and (b) Spin
polarization via anomalous magnetic moments, which increases the
pressure through magnetization of the neutron component.  The balance
between these two effects determines whether the EoS becomes softer or
stiffer in the presence of extreme magnetic fields. 

However, because the magnetic field provides additional support against gravity through
both the Lorentz force and the electromagnetic stress-energy,
self-consistent solutions of Einstein–Maxwell theory show that a star
with the same baryon number can sustain a larger gravitational mass
and radius than its unmagnetized
counterpart~\cite{Chatterjee2021Review,Thapa:2021kfo,Peterson:2023bmr}.
The magnitude of this effect
depends sensitively on the magnetic-field geometry and the underlying
EoS: it is most pronounced for globally ordered (with significant
poloidal) field configurations. 

The current understanding of the stability and internal structure of magnetars suggests that mixed magnetic-field configurations, often referred to as {\it twisted torus} geometries, can provide long-term stability over secular timescales of millions of years, comparable to the active lifetimes over which neutron stars sustain their strong magnetic fields. 

Purely poloidal or purely toroidal field configurations are known to be unstable to various magnetohydrodynamic instabilities (e.g., kink or Tayler instabilities), whereas a combined configuration can achieve a stable equilibrium by mutually stabilizing these components. In a twisted torus configuration, the poloidal field threads the star and extends smoothly into the exterior, determining the large-scale dipolar structure observed electromagnetically, while the toroidal component is predominantly confined to closed-field-line regions in the stellar interior, where it can reach strengths exceeding that of the poloidal field, see Fig.~\ref{fig:TwistedTorus}. 
Such configurations are supported by numerical solutions of the coupled Einstein–Maxwell equations for magnetized, self-gravitating fluids, and represent equilibria consistent with both relativistic gravity and realistic equations of state~\cite{Pili2014,Chatterjee2021Review}. The presence of a substantial internal toroidal field also has important implications for the star’s deformation, stability, and potential gravitational-wave emission. 

More generally, the internal magnetic-field configuration of neutron stars is still not well constrained. While global models (e.g., mixed poloidal–toroidal field configurations) exist, they are subject to uncertainties related to stability, composition, superconductivity, and coupling between different components. Direct observational constraints probe the external dipolar field primarily, whereas the internal field strength and geometry—particularly in the core—remain largely uncertain. Consequently, magnetic-field effects can be incorporated at a qualitative or phenomenological level, but a fully self-consistent microphysical and macroscopic description is still an open problem.

\begin{figure}[t]
\centering
\includegraphics[width = 0.45\textwidth]{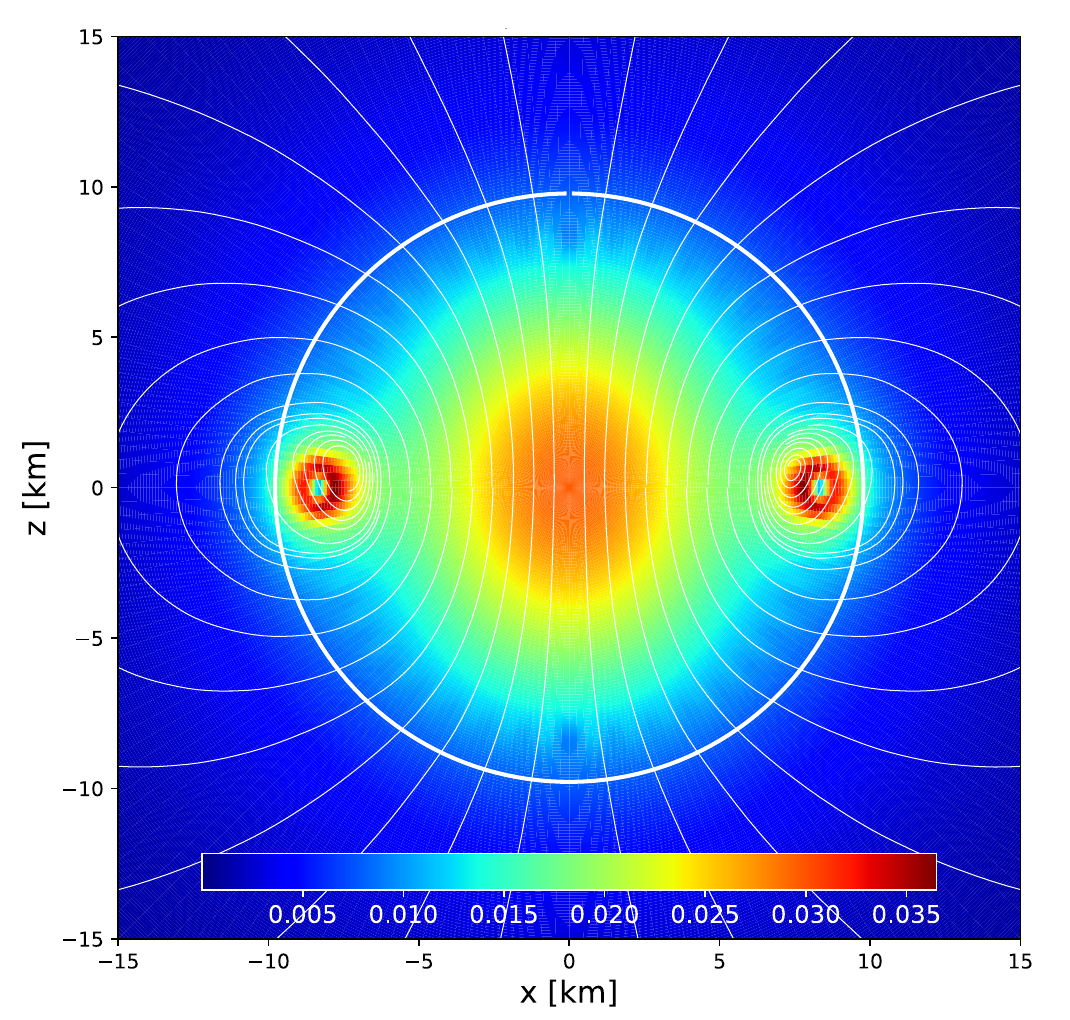}
\caption{ {\bf Twisted-torus magnetic field structure in a magnetar
    showing coexisting poloidal and toroidal components.}  Spatial
  structure of a twisted-torus magnetic field, illustrating the
  coexistence of poloidal and toroidal components in a magnetar
  obtained with the XNS solver~\cite{Das:2025fws}. The color bar shows
  the magnetic field strength in units of $10^{17}$~Gauss. The white
  lines trace the dipolar (poloidal) field component, while the
  toroidal component is confined to the darker equatorial regions.  }
\label{fig:TwistedTorus}
\end{figure}

At high densities, spin polarization effects can affect the phase
structure of the baryonic matter. An extreme, but uncertain
case, is the ferromagnetic phase transition, where the interaction
causes a spontaneous transition to a polarized state even in the absence
of the magnetic field. Both qualitative and quantitative evidence show
that the threshold densities for ferromagnetic or highly spin-polarized phases
lie generally at several times nuclear saturation
density. The precise threshold depends on the nuclear
interaction model, external field, and temperature, with some models
showing no evidence for such transition~\cite{Clark1969,Tews2020ApJ}.
Furthermore, hyperons/quark matter appear at high densities and could
preempt ferromagnetic phases. Also, while pure neutron matter may or
may not show ferromagnetic behaviour, stellar matter must be subject to
electric charge neutrality, $\beta$-equilibrium~\cite{Vidana2002},
and one needs to take into account magnetic screening or shielding by
superconductivity~\cite{Haensel1996}, all of which may alter viability
of the ferromagnetic phase.

Landau quantization of charged particles in a strong magnetic
field affects several key microphysical processes in neutron stars. In
the core, it modifies neutrino emissivity and shifts the threshold
density for the direct Urca reaction~\cite{Arras:1998mv,Maruyama2022} and
introduces Shubnikov--de Haas oscillations in transport
coefficients~\cite{Huang2010,Potekhin:2015qsa,Shovkovy:2025yvn}.
While in the core of a neutron star, where multiple
fermion species are Landau quantized, the EoS and other observables are not strongly altered for a physically relevant field strengths
$B\le 10^{18}$~G~\cite{Broderick2000}, in the less-dense crust, these quantization effects are significant. This is especially true within the outer envelope with densities $\rho\lesssim 10^{10}$ g/cm$^3$, where quantization effects enhance the thermal conductivity parallel to the magnetic field. This generally results in a higher effective surface temperature for a given envelope-crust boundary temperature for typical mixed dipole-toroidal field configurations~\cite{Ventura2001,Potekhin:2015qsa}.

Landau quantization can
alter the equilibrium nuclear composition and shift the neutron drip
line to higher or lower densities, depending on the field
strength~\cite{Chamel2015}. Furthermore, the quantization of electrons
and other charged fermions gives rise to de Haas--van Alphen
oscillations in the differential magnetic susceptibility of
matter. These oscillations can, in turn, induce the formation of
diamagnetic domains, introducing additional anisotropies in the crustal
structure~\cite{Blandford1982,Suh2010,Rau2023}. 
They can also enhance the dissipation of the magnetic field through the generation of small-scale field features of size 
\bea
\ell \sim eB \hbar c \,\mu_e \left(\frac{dz}{d\mu_e}\right)
\sim \frac{230}{g_{14}\mu_{e10}} \left(\frac{B}{10^{14}\textrm{G}}\right)\,\mathrm{cm}
\eea
for electron chemical potential $\mu_{e10}$ in units of 10~MeV, and gravitational acceleration $g_{14}$ in units of $10^{14}$ cm s$^{-2}$~\cite{Rau2025}. This is generally much smaller than the large-scale field with length scale of the order of the stellar radius $R\sim 10-13$ km or crust thickness $\sim 0.1R$.


\section{Spin and superfluidity in neutron stars}
\label{sec:Superfluidity}

\subsection{Pairing channels}
\label{sec:SuperfluidityA}

Microscopically, spin determines the pairing patterns of fermions,
leading to superfluidity and superconductivity within the stellar
interior~(see Ref.~\cite{Sedrakian2019} and references
therein). Figure~\ref{fig:gaps} shows the typical arrangement of
superfluid/superconducting phases in the interior of a neutron star
for a given composition. Neutrons in the crust and outer core are
expected to pair predominantly in the spin-singlet $ ^1S_0$ channel,
while in the higher-density inner core they form pairs in the
spin-triplet $^3P_2$ channel. Protons, which are present at the
few-percent level, pair mainly in the $^1S_0$ state and form
 type-II at low densities and type-I at high densities superconducting
condensates. The interplay of these spin-dependent condensates governs
glitch phenomena, vortex–flux tube interactions, neutrino
emissivities, and the overall thermal and rotational evolution of
neutron stars. Moreover, strong magnetic field restructures the phase
diagram of superfluids and superconductors: proton superconductivity
may be suppressed at fields above the critical value, while the
neutron $^3P_2$–$^{3}F_2$ superfluid phases can acquire spin-aligned order
parameters. These effects propagate into transport processes,
including thermal and electrical conductivities, viscosities, and
neutrino emission rates, all of which are sensitive to spin alignment
and field strength. The spin–magnetic interaction thus emerges as a
critical ingredient in the microphysics of magnetars and in
determining their observational signatures~\cite{Adhikari2026PrPNP}.

The role of spin in pairing in nuclear and neutron-star matter can be
understood from the partial-wave analysis of nucleon-nucleon ($NN$)
scattering. Two-nucleon states are labeled using standard
spectroscopic notation ${}^{2S+1}L_J$, where $L = 0,1,2,\dots$
corresponds to $S, P, D, \dots$ waves, $S = 0,1$ indicates singlet or
triplet spin states, and $J$ is the total angular momentum.  The Pauli
principle requires that the total wave function, including spin and
isospin, be antisymmetric, which imposes the condition
$L + S + {\sf T}$ odd.

At low energies, $L=0$ states dominate, giving rise to the $^{1}S_0$ and
$^{3}S_1$–$^{3}D_1$ coupled channels. In neutron-rich matter, where
neutrons dominate, only ${\sf T}=1$ pairing is allowed, so the
attractive $^{3}S_1$–$^{3}D_1$ channel is forbidden.  In contrast,
symmetric nuclear matter can support ${\sf T}=0$ pairing in this
channel. At asymptotically low densities, this may lead to a
transition to a Bose-Einstein condensate of deuterons; however, 
higher-order clustering effects are also expected.

Theoretical calculations of pairing are largely based on
phase-shift–equivalent nucleon–nucleon potentials, such that the
dominance of a given channel is constrained by experimental scattering
phase shifts and, in the case of the $^{3}S_1$–$^{3}D_1$ channel, by
the existence of the deuteron bound state. However, the relative
strength of different pairing channels can be significantly modified
by many-body effects, which may either enhance or reduce the effective
attraction and thus alter the resulting pairing gap. In particular,
medium polarization effects are known to play an important role, yet
their quantitative impact remains uncertain, and no clear consensus
has been reached regarding their magnitude~\cite{Sedrakian2019}.

In neutron-star matter, with neutron fractions around $95\%$ at
saturation and gradually decreasing at higher densities, ${\sf T}=1$
pairing dominates. Neutron pairing occurs in the $^{1}S_0$ channel up
to slightly above the crust-core interface, while proton pairing in the same channel persists to higher densities due to their lower Fermi
energies. At higher energies, the $^{3}P_2$–$^{3}F_2$ coupled partial
wave becomes the dominant ${\sf T}=1$
channel, while at lower-energies (and, thus, densities)
$^{3}P_0$ and $^{3}P_1$ channels are subdominant or repulsive.
\begin{figure}[t]
\centering
\includegraphics[width = 0.5\textwidth]{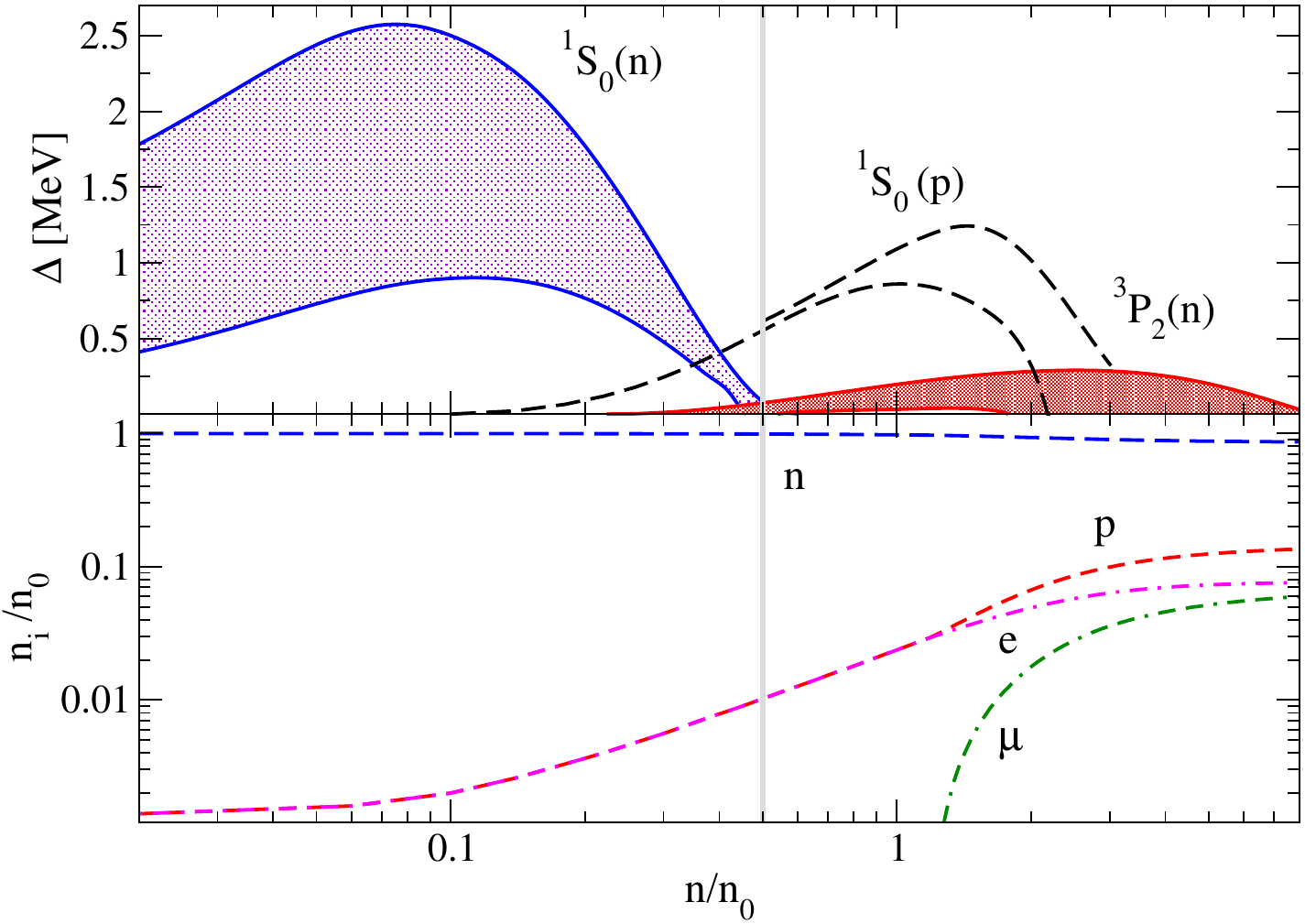}
\caption{
{\bf Density dependence of neutron and proton pairing gaps and corresponding particle composition in the neutron star core.}
  {Upper panel:} Pairing gaps as functions of baryon density
  (in units of nuclear saturation density $n_0$) for neutrons in the
  $^1S_0$ (solid lines) and $^3P_2$–$^3F_2$ (dash–dotted lines)
  channels, and for protons in the $^1S_0$ channel (dashed lines). The
  shaded areas show the upper and lower bounds on neutron gaps; the
  upper and lower dashed lines show the same for proton $S$-wave gap,
  see Ref.~\cite{Sedrakian2019} and references therein. {Lower panel:}
  Particle composition of the neutron star core at $T = 0$,
  corresponding to the conditions used in computing the pairing
  gaps. The vertical line marks the crust–core interface at
  $n/n_0 = 0.5$.  }
\label{fig:gaps}
\end{figure}

\subsection{Spin-1 pairing}
\label{sec:SuperfluidityB}

The $^3P_2$–$^3F_2$ pairing in neutron matter is of major
phenomenological importance, since the neutron fluid fills most of the
stellar core. This form of triplet, odd-parity pairing introduces
several new features compared with $^1S_0$ and $^3S_1$–$^3D_1$
pairing~\cite{Zverev2003}. Different
magnetic substates can compete, while strong spin–orbit and tensor
couplings mix the $^3P_2$ and $^3F_2$ waves.

The microscopic treatment begins with a partial-wave expansion of the
pairing interaction and a decomposition of the gap into
angular-momentum components. Because the gap equation is nonlinear,
these components are coupled. A common simplification is to average
over angles, reducing the problem to a one-dimensional integral
equation. In most cases, a single channel dominates, but tensor forces
couple the $P$- and $F$-waves, leading to a system of coupled
equations similar to that found in the $^3S_1$–$^3D_1$ channel at low
densities and near isospin-symmetric nuclear matter.  Quantitative
predictions for the $^3P_2$–$^3F_2$ pairing gap depend sensitively on
three-nucleon (3N) forces, which become increasingly important as the
density increases. Calculations using the Argonne $V_{18}$ potential
with the Urbana UIX 3N interaction~\cite{ZuoCuiLombardoSchulze2008} or
the Bonn-B interaction, including 3N forces~\cite{DongLombardZuo2013},
typically yield a maximum gap of about 0.5 MeV, see
Fig.~\ref{fig:gaps}. However, beyond mean-field
Bardeen-Cooper-Schrieffer (BCS) theory, effects such as wave-function
renormalization can reduce this value by an order of magnitude, see
Ref.~\cite{Ding2016}. The gap is also strongly sensitive to the choice
of 3N interaction and emphasizes the need for consistency between
two-and three-nucleon
forces~\cite{Papakonstantinou:2017ewy}. Importantly, neutron
$^3P_2$–$^{3}F_2$ superfluidity is of broad and interdisciplinary
relevance, as it provides a concrete link between the microphysics of
dense nuclear matter in neutron stars and a wider class of systems
exhibiting topological quantum order. In particular, its anisotropic,
spin-triplet pairing and associated symmetry-breaking patterns closely
parallel those realized in superfluid $^3$He, unconventional (e.g.,
heavy-fermion) superconductors, and engineered ultracold atomic gases,
making it a valuable platform for exploring topological phases,
collective excitations, and quantum vortices in different areas of
physics.

At even higher densities, isospin-symmetric nuclear matter favors
$^3D_2$ pairing, but increasing neutron–proton asymmetry drives a
transition to $^3P_2$–$^3F_2$ pairing. Because even a small isospin
imbalance destroys $D$-wave pairing~\cite{AlmSedrakian1996}, this
state can exist only if the deep cores of neutron stars contain nearly
symmetric matter, for example, when $K^-$ condensation sets in.

\subsection{Suppression of pairing by magnetic field}
\label{sec:SuperfluidityC}

Physical phenomena in dense, strongly interacting matter that depend
on the quasiparticle spectrum near the Fermi surface are influenced by
magnetic fields of order $B \sim 10^{16}-10^{17}$ G, which are several
orders smaller than those required to affect the EoS, as argued in
Sec.~\ref{sec:Bfields}. This is particularly relevant for nucleonic
pairing and for neutrino radiation processes, which dominate the
cooling of compact stars. The field interacts with neutron or proton
spins, creating an imbalance between spin-up and spin-down particles,
which suppresses Cooper pairing since not all spin-up particles have
available spin-down partners~\cite{Stein2016}. This Pauli paramagnetic
suppression affects both neutron and proton condensates, but is the
dominant mechanism for neutrons. Proton pairing, in contrast, is more
sensitive to weaker fields through a different mechanism: the Larmor
motion of charged protons in the magnetic field, arising from their
charge–field interaction~\cite{SinhaSedrakian2015,Haber2017}.

BCS superconductors are characterized by
three length scales: (i) the London penetration depth $ \delta $,
(ii) the coherence length $ \xi $, and (iii) the interparticle
distance $ d $. In neutron stars, the condensates satisfy the condition
$ d \ll \delta, \xi $, corresponding to the weak-coupling BCS
regime. The ratio of the first two scales defines the
Ginzburg-Landau parameter $ \kappa_{\rm GL} = \delta/\xi $
\cite{TinkhamBook}, which determines the type of superconductivity:
$ \kappa_{\rm GL}  < 1/\sqrt{2} $ corresponds to type-I, while
$ \kappa _{\rm GL} > 1/\sqrt{2} $ indicates type-II.

In type-II superconductors, magnetic flux penetrates via quantized
vortices carrying flux $ \phi_0 = \pi\hbar c/e $, whereas in type-I
superconductors, the field is either expelled or the matter is split
into macroscopic domains of normal and superconducting domains, if the
expulsion is ineffective. Combining $ \delta $ and $ \xi $ with
$ \phi_0 $ defines three characteristic magnetic fields:
$$
H_{c1} \simeq \frac{\phi_0}{\delta^2}, \quad
H_{cm} \simeq \frac{\phi_0}{\xi \delta}, \quad
H_{c2} \simeq \frac{\phi_0}{\xi^2}.
$$
For type-II superconductors ($ \kappa_{GL}  \ge 1/\sqrt{2} $), fluxtubes
exist in the field regime $ H_{c1} \le H_{cm} \le H_{c2} $:
specifically, above $H_{c1}$, the formation of a single flux tube
(Abrikosov vortex) becomes energetically favorable; at $H_{cm}$ -- the
thermodynamic critical field -- the energy densities of the
superconducting and normal states are equal; finally, $ H_{c2} $
corresponds to the field at which superconductivity vanishes due to
the overlap of vortex cores; thus, for $H>H_{c2}$, superconductivity
of protons is destroyed.  

The suppression of pairing in the
$^1S_0$-wave neutron condensate differs from that in the proton
condensate because charge-neutral neutrons interact with the $B$-field
only via their spin magnetic moment (Pauli paramagnetism). The
corresponding critical field for destruction of pairing by
spin-polarization is by an order of magnitude
larger~\cite{Stein2016}. In the neutron star's core, neutron pairing
shifts to the $^3P_2$–$^3F_2$ channel, which involves total spin-1
pairs; in this case, the magnetic field does not destructively affect
the internal spin structure~\cite{Muzikar1980,Sauls1981,Sauls1982}. As
density increases and the proton gap is diminished, their coherence
length increases, and eventually the condition for proton type-I
superconductivity $ \kappa_{GL}  \le 1/\sqrt{2} $ is satisfied. For fields $H> H_{cm}$, the matter is in the normal state; otherwise, a (local) Meissner expulsion of the field and break-up into normal-superconducting domains occurs.

Quite generally, the absence or
reduction of neutron superfluidity affects many crustal microphysical
properties, including neutrino emissivity, transport, thermal
relaxation, and dynamical coupling times, which in turn influence the
damping of oscillations and the interpretation of glitches and
anti-glitches. While Pauli paramagnetism also applies to
$S$-wave proton pairs, proton pairing is suppressed by the diamagnetic
effect associated with the curvature of the proton trajectories in a
magnetic field at lower fields, as discussed above, 
see also Ref.~\cite{Adhikari2026PrPNP}

\begin{figure}[t]
\centering
\includegraphics[width = 0.5\textwidth]{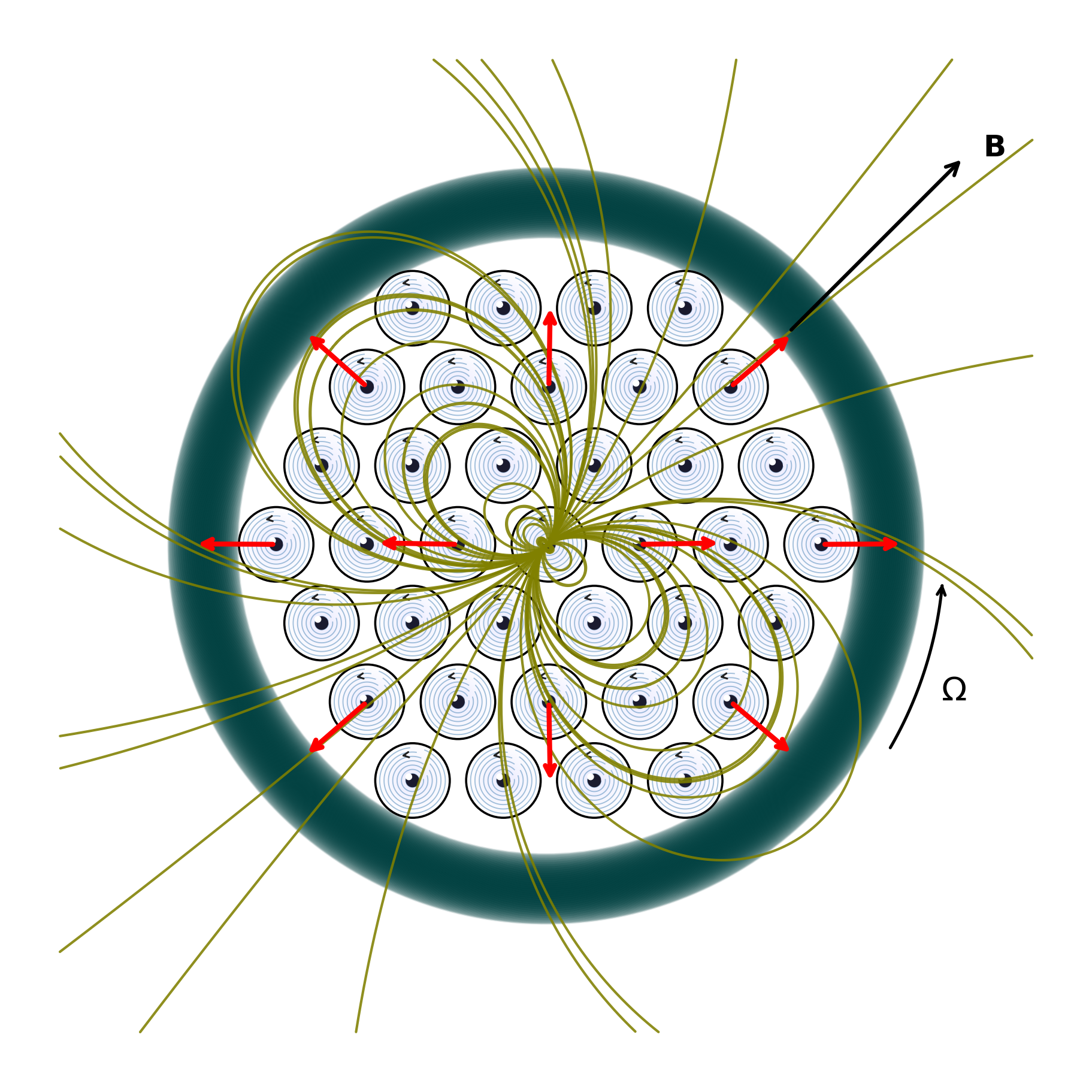}
\caption{ {\bf Neutron star cross section illustrating superfluid
    vortex lattice, magnetic flux tubes, and spindown–driven vortex
    motion.}  A cross-section of a neutron star perpendicular to the
  spin axis, illustrating the triangular lattice of neutron vortices
  (shown not to scale) that carries the angular momentum of the
  neutron superfluid. The dipolar magnetic field is also shown, with
  its axis aligned along the tilted $\vecB$-vector, together with the
  corresponding field lines, which in superconducting proton regions
  are confined to flux tubes. The red arrows indicate the radial
  outward motion of neutron vortices in response to the star’s secular
  spindown.  }
\label{fig:vortex_lattice}
\end{figure}

\subsection{Quantum vorticity and mutual friction}
\label{sec:SuperfluidityD}

The study of vorticity in nuclear systems is motivated by the rotation
of neutron stars, where neutrons form a neutral superfluid that
rotates via an array of quantized vortices~\cite{Sauls2019}. 
While vortex states in finite nuclei have been conjectured, the condensate coherence length is comparable to or larger than nuclear radii, and one cannot speak of vortices in the usual sense. In contrast, vorticity in neutron stars shares many features with laboratory superfluids and superconductors, including Bose-condensed $^4$He, fermionic $^3$He, and ultracold atomic gases~\cite{PitaevskiiBEC}.

In neutron stars, rotation at angular velocity $\Omega$ generates
neutron vortices with density
\bea\label{eq:Feynman}
n_n^{(V)} = \frac{2\Omega}{\kappa}, \qquad \kappa = \frac{\pi \hbar}{m^*_n},
\eea
where $m^*_n$ is the neutron effective mass, 
while type-II proton superconductivity produces electromagnetic vortices with density
\bea
n_p^{(V)} = \frac{B}{\phi_0}.
\eea
In general, the rotation and magnetic axes are misaligned, as illustrated in Fig.~\ref{fig:vortex_lattice}. As the star spins down due to magnetic dipole radiation braking, its rotation frequency decreases. According to Eq.~\eqref{eq:Feynman}, this implies a reduction in the neutron vortex density, which is achieved through an outward radial expansion of the vortex lattice (along the cylindrical radius), as indicated by the arrows in Fig.~\ref{fig:vortex_lattice}. The corresponding radial velocity of the vortices is
\bea
v_r = -\left(\frac{\dot \Omega}{2\Omega } \right)r,
\eea
which follows from the conservation of the vortex number density $n_V$.

Both neutron vortex and proton fluxtube lattices are triangular, with
basis lengths
\bea
d_n = \left(\frac{\kappa}{\sqrt{3}\, \Omega}\right)^{1/2}, \quad
d_p = \left(\frac{2\, \phi_0}{\sqrt{3}\, B}\right)^{1/2},
\eea
which are of order $10^{-4}$ cm and $10^{-9}$ cm, respectively, for typical neutron-star parameters.  These lengths define a mesoscopic scale for neutron-star superfluids, bridging the microscopic scale of the vortex core, set by the coherence length $\xi$ (where the
superfluid order parameter vanishes) and the macroscopic scale of the stellar core or crust.  

The fermionic vortex core hosts quasiparticle
bound states described by Bogolyubov–De Gennes
theory~\cite{Gennes1999superconductivity}. While in
the $^1S_0$ condensate, neutron vortices are well-understood, the situation is more complex in the $P$-wave superfluid, which may
exhibit non-trivial topological properties in analogy to the liquid
$^3$He in the condensed matter context. For example, Weyl and Majorana
fermions can emerge as quasiparticle excitations in neutron $^3P_2$
superfluids~\cite{Masaki:2021hmk}. In addition, there are several
branches of bosonic collective modes associated with the spin, in
addition to the standard phonon and Higgs modes in $S$-wave superfluid
~\cite{Bedaque:2012bs,Bedaque:2013fja,Leinson:2010pk}.

It was recognized early on that quantum vortices in neutron $^3P_2$
superfluids can exhibit spontaneous
magnetization~\cite{Muzikar1980,Sauls1981,Sauls1982}.  More recent
studies suggest that these vortices may also host Majorana fermions or
split into half-quantized non-Abelian
vortices~\cite{Masaki:2021hmk,Kobayashi:2022moc,Masaki:2023rtn}. At
the interface between the $^3P_2$ and $^1S_0$ superfluid phases,
topological surface defects may
form~\cite{Yasui:2019pgb}. Furthermore, because the phases of neutron
$S$- and $P$-wave condensates are different on both sides of their
interface, it acts as a Josephson junction, which can excite plasma
waves under non-stationary conditions~\cite{Sedrakian_Rau2025}. The
subsequent dissipation of these waves can lead to heating.

Another distinctive feature of the neutron $^3P_2$ phase is its
coexistence with the proton $^1S_0$ superconductor. These two
condensates are entrained, meaning that the motion of one induces
motion of the other. In hydrodynamic terms, their velocities and
momenta are linked through a two-by-two entrainment matrix with
off-diagonal elements coupling the neutron and proton superflows. A
key phenomenological consequence of this entrainment is the appearance
of a non-quantized magnetization on neutron
vortices~\cite{Sedrakian1980,Alpar1984ApJ}—one that exceeds their
intrinsic spin-induced magnetization—and thereby modifies their
coupling to the normal electron and muon fluids~\cite{Alpar1984ApJ}.

Mutual friction arises from the interaction of vortices with the ambient non-superfluid components of neutron star matter and is central to understanding neutron-star rotational dynamics, including glitches and postglitch relaxation~\cite{Sauls2019,Haskell:2017lkl,Antonopoulou2022}. Analogous dissipative mechanisms are well known in other quantum fluids, including both bosonic superfluid $^4$He and fermionic superfluid $^3$He, as well as in ultracold atomic Fermi gases. Neutron stars are nevertheless unique in that they involve the interplay of multiple condensates—most notably a neutron superfluid coexisting with a charged proton superconductor—coupled to a relativistic, highly degenerate electron background under extreme conditions of density and gravity.

For example, electrons scatter off neutron vortex-core quasiparticles
via the coupling of their charge $-e$ to the neutron magnetic moment. The electron
relaxation time in an $S$-wave neutron superfluid is~\cite{Bildsten1989}
\bea
\tau_{eV}[^1S_0] = \frac{1.6\times 10^3}{\Omega} 
\frac{\epsilon_{Fe}^2\Delta}{\epsilon_{Fn}^2T}
\left(\frac{\epsilon_{Fn}}{2 m_n}\right)^{1/2}
\exp\left(\frac{\epsilon^0_{1/2}}{T}\right),\nonumber\\
\eea
where $\Delta$ is the pairing gap, $\epsilon_{Fi}$ are Fermi energies of electrons $(e)$ and $(n)$,  
$\epsilon^0_{1/2}$ is the lowest vortex-core state, and the angular
velocity of rotation $\Omega$ enters via the vortex density $n_n^{(V)}$. The collision rate $\tau_{eV}^{-1}$ is seen to be exponentially suppressed at low $T$.

For a $P$-wave neutron superfluid, the order parameter is a traceless
tensor $A_{\mu\nu}$,
\bea
A_{\mu\nu} &=& \frac{\Delta}{\sqrt{2}} e^{i\phi} \big[ f_1 \hat r_\mu
\hat r_\nu
+ f_2 \hat \phi_\mu \hat \phi_\nu \nonumber\\
&-& (f_1+f_2) \hat z_\mu \hat z_\nu + i f_3 (r_\mu \hat \phi_\nu + r_\nu \hat \phi_\mu) \big],
\eea
where $f_{1,2,3}(r)$ describe the radial
profile~\cite{Sauls1982}. $P$-wave vortices are intrinsically
magnetized due to spin-1 Cooper pairs, and electron scattering off
them yields a temperature-independent relaxation time~\cite{Sauls1982}
\bea
\tau_{eV}[^3P_2] \simeq \frac{7.91\times 10^8}{\Omega}
\left(\frac{k_{Fn}}{\text{fm}^{-1}}\right) \left(\frac{\text{MeV}}{\Delta}\right) \left(\frac{n_e}{n_n}\right)^{2/3}.
\eea
where $n_i$ are the number densities of species, $k_{Fn}$ is the
neutron Fermi wave-vector.  Furthermore, the entrainment of proton
supercurrent by the neutron vortex circulation induces an effective
vortex flux
\bea
\phi^* = k_{\rm ent} \phi_0, 
\eea
where $k_{\rm ent} $ depends on the proton quasiparticle mass,
enhancing the vortex
magnetization~\cite{Sedrakian1980,Alpar1984ApJ}. The associated
electron relaxation time is~\cite{Alpar1984ApJ} 
\bea
\tau^{-1}_{e\phi} = \frac{3\pi}{32}
\left(\frac{\epsilon_{Fe}}{m_pc^2}\right)
\frac{\tau_0^{-1}}{k_{eF} \delta}, \quad
\tau_0^{-1} = \frac{2 cn_n^{(V)}}{k_{eF}}
\left( \frac{\pi^2 \phi_*^2}{4 \phi_0^2} \right),
\eea
where $k_{eF}$ is electron Fermi wavenumber, $m_p$ is the proton
mass. This relaxation depends weakly on temperature, reflecting its
electromagnetic origin.  A full treatment of mutual friction also
involves interactions between neutron and proton vortices. If co-axial
proton clusters are formed around neutron vortices, the coupling to
the electron fluid may increase by orders of
magnitude~\cite{Sedrakian1995}.


\section{Superfluidity and astrophysics of compact stars}
\label{sec:Astro}

\subsection{Glitch behavior  and recovery}

Pulsar glitches are sudden jumps in their rotation rate and its
derivative that interrupt the otherwise gradual spindown of pulsars,
first observed in the Vela pulsar PSR B0833–45. In the 2016 Vela
event, the glitch rise was resolved on a pulse-to-pulse basis and
constrained to last $\tau_r \lesssim 16~\mathrm{s}$. The subsequent
relaxation of both the rotation frequency and the spindown rate
($\dot\nu$), occurring over timescales from minutes to months, points
to the presence of a loosely coupled superfluid component within the
star~\cite{Baym1969Natur}. A representative example of such a glitch
and its postglitch evolution in the derivative of the rotation rate of
the Vela pulsar is shown in Fig.~\ref{fig:glitch}. It is seen that a
rapid (exponential) recovery phase is followed by a quasi-linear
long-term relaxation, which, in general, does not return to the
pre-glitch extrapolated (in the absence of glitch) values, showing
permanent frequency offsets.

Glitches are observed to repeat in several young pulsars, most notably in Vela and the Crab, and across the pulsar population they span a wide range of magnitudes—from small events to so-called giant glitches—highlighting the richness of the underlying physics. From a theoretical perspective, this phenomenology raises several key questions: how and where angular momentum is stored within the star, presumably in a superfluid reservoir; what mechanism enables its rapid release on such short timescales; and what cyclic process can sustain the quasi-regular recurrence of glitches.
\begin{figure}[t]
\centering
\includegraphics[width = 0.5\textwidth]{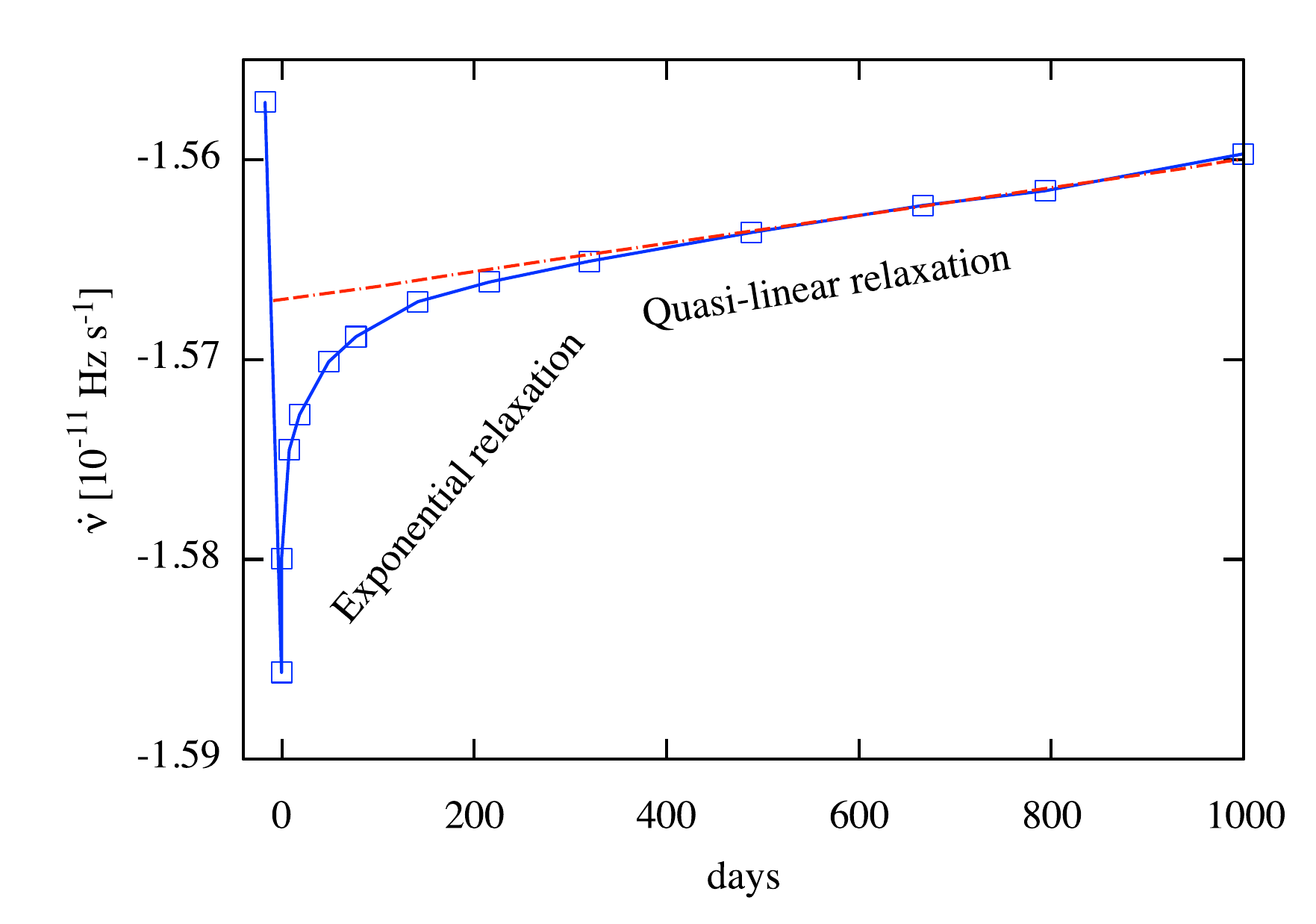}
\caption{{\bf postglitch evolution of the Vela pulsar’s spindown
    rate illustrating fast relaxation and extended recovery.}
  Illustration of the glitch and postglitch relaxation in the
  rotational frequency derivative of the Vela pulsar PSR B0833–45,
  where the origin of time corresponds to the glitch. The initial,
  quasi-instantaneous jump in rotation rate is followed by an exponential relaxation
  over timescales of days to months, and subsequently by a
  quasi-linear long-term recovery extending over several years. Data
  points are adopted from figure 5 in~\cite{Cordes1988}.  The dashed
  line shows the secular spindown derivative.}
\label{fig:glitch}
\end{figure}
The dynamics of superfluid interiors provide the most promising
framework for explaining glitch behavior. From a theoretical
perspective, this involves the motion of quantized vortices in a
neutral superfluid—either neutrons in the crust and core or,
potentially, superfluid condensates such as the color-flavor–locked
(CFL) phase in quark matter. It is now understood that these dynamics
operate in two distinct regimes, linear and nonlinear, depending on
how the lag between the rotation of the superfluid and that of the
normal component evolves under the influence of mutual friction and
vortex pinning.

In the linear regime, superfluid hydrodynamics predicts a linear
relation between the forces acting on vortices and the relative
velocities (lags) between the superfluid, the vortices, and the normal
component. In particular, the Magnus force depends on the lag between
the superfluid and vortex velocities, while the mutual friction force
depends on the lag between the vortex and normal component
velocities. Restricting ourselves to the Newtonian case for clarity,
the corresponding force balance equation on the vortices can be
written as~\cite{BekarevichKhalatnikov1961}
\begin{eqnarray}\label{eq:force_balance}
\rho_S\left[\left({\vecv}_S
+\vec\nabla\times(\lambda_V\vecvarpi)-{\vecv}_L\right)\times
\veckappa\right]
- \eta\left({\vecv}_L-{\vecv}_N\right)
+\eta'\left[\left({\vecv}_L-{\vecv}_N\right)\times\vecvarpi\right] =
0,
\end{eqnarray}
where $\rho_S$ and $\vecv_S$ denote the superfluid density and velocity, $\vecv_N$ and $\vecv_L$ are the velocities of the normal component and vortices, respectively, and $\veckappa$ characterizes the vortex circulation with $\vecvarpi = \veckappa/\kappa$, $\lambda_V$
is the vortex tension and $\eta'$-term is the transverse friction (Iordanskii) force. 

Within the hydrodynamic framework, one can solve for the response of the normal component (the crust together with the charged plasma) to a sudden perturbation. The resulting evolution of the observable rotation frequency and its derivative in a multi-superfluid shell star is given 
by~\cite{Cordes1988,SedrakianSedrakian1995}
\begin{eqnarray}
\label{eq:nu}
\nu(t) &=& \nu_0-\frac{\nu_0}{\tau_0}t\nonumber\\
&-&\sum_i\frac{I_{Si}}{I_c}
\left(\frac{\nu_0}{\tau_0}\tau_i-\Delta\nu_{Si}\right)
\left(1-e^{-t/\tau_i}\right), \\
\label{eq:nudot}
\dot\nu(t) &=& -\frac{\nu_0}{\tau_0}-\sum_i\frac{I_{Si}}{I_c}
\left(\frac{\nu_0}{\tau_0}\tau_i-\Delta\nu_{Si}\right)
\frac{e^{-t/\tau_i}}{\tau_i},
\end{eqnarray}
where $\nu_0$ is the pulsar’s rotation frequency at the reference time (typically just before the glitch) and $\tau_0$	
  is the characteristic spindown timescale, defined through the external torque acting on the star. Fitting these expressions to observational data allows one to extract the characteristic coupling timescales $\tau_i$ and the associated superfluid moments of inertia $I_{Si}$, given the moment of inertia of the normal component $I_c$, which to a good approximation, is equal to the total moment of inertia of the star.

The relaxation timescale appearing in Eqs.~\eqref{eq:nu} and \eqref{eq:nudot}, which characterizes the exponential decay of the postglitch response, can be related to the parameters entering the force balance equation~\eqref{eq:force_balance} 
as~\cite{AlparSauls1988,Sedrakian1995}
\bea\label{eq:tau_d}
\tau=\frac{1}{2 \Omega_s(0)}\left( \frac{\eta}{{\rho}_S \kappa}
+\frac{{\rho}_S \kappa}{\eta}\right),
\eea
where, for simplicity, we have suppressed the index $i$ labeling different superfluid regions. Here $\Omega_s(0)$ is the initial angular velocity of the superfluid, $\eta$ denotes the mutual friction coefficient describing the interaction between vortices and the normal component (predominantly ultrarelativistic electrons). In writing this expression, we have neglected the transverse friction term by setting $\eta' = 0$.  Notably, the dynamical relaxation time has a non-trivial dependence on the viscous
friction coefficient $\eta$. It takes on large values in two limiting cases $\eta \ll {\rho}_S \kappa$ and $\eta \gg {\rho}_S \kappa$,
corresponding to the weak- and strong-coupling limits between the vortices and the normal component. The minimum of $\tau_d$ is
achieved, when $\eta={\rho}_S \kappa$ with
$\tau_d(\min )=\Omega_s^{-1}(0)$, which sets the absolute minimum on
the relaxation time.

\begin{figure}[t]
\centering
\includegraphics[width = 0.5\textwidth]{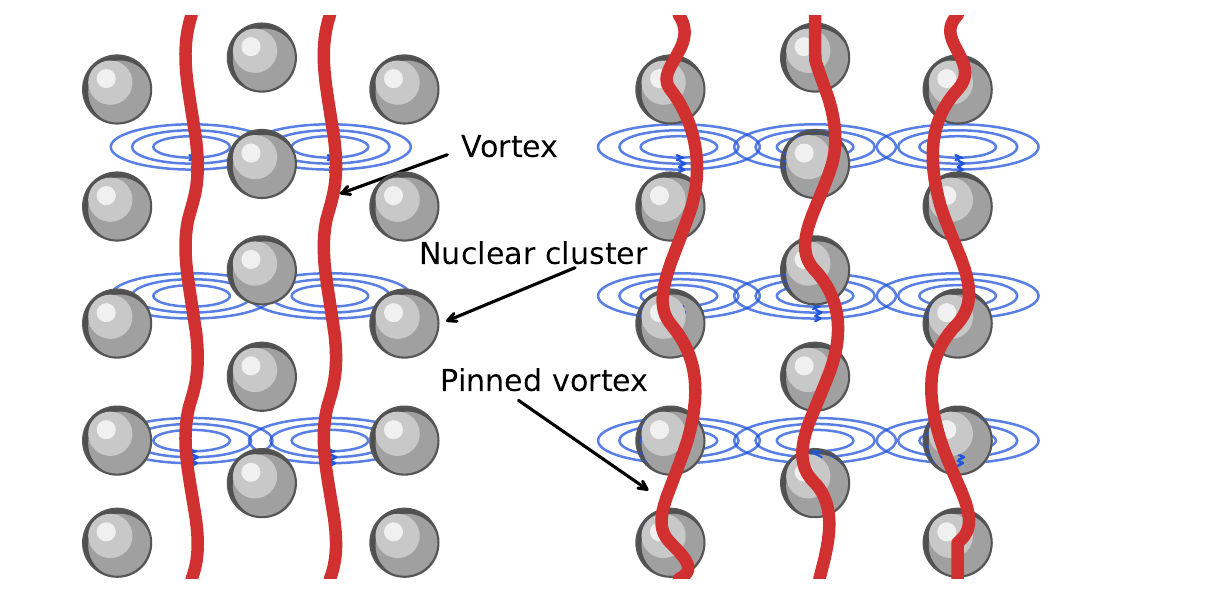}
\caption{
{\bf Two regimes of vortex dynamics in the neutron star crust: free
  motion and thermally activated creep of pinned
  vortices.}
  Illustration of two distinct regimes of vortex dynamics in the neutron star crust: on the left, vortices move freely through the nuclear lattice (the unpinned, linear regime), while on the right, vortices are pinned to lattice nuclei. In the latter case, vortex creep proceeds via thermally activated, jump-like motion of vortices from one pinning center to another.
 }
\label{fig:vortex_dynamics}
\end{figure}

Although the weak-coupling regime was initially invoked to explain
glitches—attributing the long relaxation times to weak coupling
between vortices and the electron fluid~\cite{Baym1969Natur}—an
alternative picture emerged in the 1970s, inspired by the resistive
state of type-II superconductors. In that context, for small driving
currents, a vortex lattice responds via thermally activated motion,
leading to an exponential creep of vortices through pinning centers.

Neutron star crusts host both a lattice of quantized vortices and a
lattice of nuclear clusters, which coexist and interact with each
other. The nuclear clusters can act as pinning centers, since certain
configurations of a vortex relative to a cluster minimize the total
energy of the system. This leads to the possibility of vortex pinning,
where vortices become temporarily anchored to the crustal
lattice. Fig.~\ref{fig:vortex_dynamics} illustrates two limiting
regimes: on the left, vortices move freely under the action of forces
in the linear (unpinned) regime, while on the right, vortices are
pinned to the nuclear lattice.

P. W. Anderson and N. Itoh~\cite{Anderson:1975zze}
proposed that glitches arise from the sudden unpinning of a superfluid
component that is otherwise decoupled from the observable crust,
resulting in a rapid transfer of angular momentum. In more detail,
before a glitch, the crust spins down under the action of an external
torque, while the superfluid component remains partially decoupled,
leading to a gradual buildup of a rotational lag $\Delta \Omega$
between the two components. When this lag exceeds a critical
threshold, $\Delta \Omega \gtrsim \Delta \Omega_c$, vortices that were
previously pinned to the crust suddenly unpin. This triggers the
glitch itself, which occurs on very short timescales (seconds or less)
and involves a rapid transfer of angular momentum from the superfluid
to the crust, resulting in $\Omega_c$ increasing and $\Omega_s$
decreasing. Following the glitch, the system undergoes a relaxation
phase over timescales ranging from minutes to months, during which the
superfluid and normal components gradually re-couple through mutual
friction or vortex creep. This recovery is typically well described by
exponential relaxation with characteristic timescales $\tau_i$.

In crustal vortex creep models, the postglitch relaxation is governed by thermally activated vortex motion (see Ref.~\cite{Link2014} and references therein), with a radial velocity
\bea
v_r \approx v_0 \exp(-E_a/k_B T),
\eea
where $v_0 \sim 10^7$ cm s$^{-1}$, and the activation energy is given by $E_a \approx E_p \big(1-\Delta\Omega/\Delta\Omega_c\big)$. Here $\Delta\Omega$ denotes the lag between the superfluid and the crust, $\Delta\Omega_c$ is the critical lag for unpinning, and $E_p$ is the pinning energy. The evolution of the observable crustal angular velocity $\Omega_c$ then follows
\bea
I_c \dot{\Omega}_c = N_{\rm ext} +
\sum_i I_s^i \frac{2 \Omega_s^i}{r} v_r^i,
  \eea
  where $I_c$ and $I_s^i$ are the moments of inertia of the normal component and the $i$-th superfluid component, respectively, $\Omega_s^i$ is the angular velocity of that component, $r$ is the cylindrical radius, and $N_{\rm ext}$ is the external torque.

A generic consequence of angular momentum conservation during a glitch,
\bea
I_c \Omega_c + \sum_i I_s^i \Omega_s^i = \text{const},
\eea
is that a sudden exchange of angular momentum between the superfluid interior and the solid crust satisfies  $I_c \Delta\Omega_c + \sum_i(I^i_s \Delta\Omega^i_s) = 0$. The observed amplitudes of large glitches—such as those in the Vela pulsar—require that a superfluid component with a fractional moment of inertia of order $\sum_i I_s^i / I \sim 10^{-2}$ participates in the dynamics, where $I$ is the total stellar moment of inertia. Furthermore, analyses of Vela’s glitches over several decades indicate that these large events require at least $\gtrsim 1.4\%$ of the star’s moment of inertia to be involved, thereby setting a lower bound on the superfluid reservoir~\cite{Link:1999ca}.

Although the crystalline lattice of the inner crust provides a natural
site for vortex pinning, it also introduces band-structure effects: neutrons occupy Bloch states rather than moving freely, which can
substantially reduce the effective superfluid
fraction~\cite{Chamel2013}. This finding implies that crust-only
models are insufficient, since the moment of inertia associated with the crustal superfluid is significantly smaller. More recent work
indicates that allowing interband transitions can restore the
superfluid density to values approaching those of quasi-free
neutrons~\cite{Almirante:2025cqe}. For complementary low-dimensional
analyses, see Ref.~\cite{Watanabe:2017nzj}.

Taken together, these results suggest that the superfluid component
responsible for large glitches may extend beyond the crust,
including part of the outer core, where neutron pairing in the
$^3P_2$ channel provides an additional reservoir of angular momentum
required to reproduce the observed glitch magnitudes. If the core were
devoid of proton flux tubes, the magnetization of $P$-wave neutron
vortices, induced by entrainment, would couple the core superfluid to
the normal component on timescales much shorter than those inferred
from postglitch relaxation~\cite{Alpar1984ApJ}.

However, the presence of proton flux tubes in the superconducting core can significantly modify this picture, as they may act as pinning centers for neutron vortices, thereby adding to the mutual coupling between the superfluid and the normal component~\cite{Sauls2019}. As illustrated in Fig.~\ref{fig:vortex_lattice}, the motion of neutron vortices can be hindered through their interaction with the flux tube array. To some extent, this situation is analogous to the pinning of neutron vortices to the crustal lattice, although it occurs in the core and involves magnetic structures rather than nuclear clusters. 

In fact,
core-based glitch models have been developed within the vortex–cluster
framework, in which each neutron vortex is surrounded by a coaxial,
dense mesh of proton flux tubes whenever the entrainment-induced magnetic field of the vortex exceeds the lower critical field for
flux-tube formation~\cite{Sedrakian1995}. In this regime, the electron
scattering off vortex clusters, dressing each neutron vortex leads to
strong coupling limit ($\eta \gg \rho_S\kappa$), consequently, to
long dynamical relaxation timescales consistent with observations~\cite{Sedrakian1995,Sedrakian:1998ki}.  In the presence of toroidal fields, neutron vortex creep against the equatorially distributed flux-tube arrays in the core can provide an
alternative mechanism for glitch recovery~\cite{Gugercinoglu:2014cda}.

The glitch mechanism itself may involve a combination of vortex
dynamics and crust quakes, the latter relieving stresses that
accumulate as the star decelerates. In self-organized criticality
models, avalanches of vortex unpinning can trigger glitches, producing
scale-invariant, power-law distributions of glitch sizes and
exponential waiting-time distributions—features consistent with most
pulsars except for Vela and PSR J0537–6910, which exhibit
quasi-periodic behavior~\cite{Antonopoulou2022}. Small glitches may be
suppressed when the number of freely moving vortices is limited or
when mutual friction is reduced.

Alternative or complementary
triggers include crustal fractures and hydrodynamical instabilities
within the superfluid interior and phenomena at the crust-core
interface~(for a review see Ref.~\cite{Haskell:2017lkl}). In addition,
both classical and quantum (superfluid) turbulence may play an
important role in the nonlinear dynamics of the superfluid and in the
coupling between its
components~\cite{Hossain2022}.

Poli {\it et al.} \cite{Poli2023} leveraged the analogy between
neutron stars and dipolar supersolids to gain full experimental access
to the system. In particular, changes in the moment of inertia arising
from internal dynamics can now be accurately measured in the
laboratory, providing a clear pathway to modeling neutron star
glitches in controlled experimental settings. This approach opens new
opportunities to investigate the microscopic mechanisms underlying
glitches and to directly connect observed dynamical signatures to
internal vortex and lattice dynamics.

Marmorini {\it et al.}~\cite{Marmorini:2020zfp} attribute neutron star
glitches to quantum vortex networks formed at the interface between
two superfluid phases in the core: a $P$-wave neutron superfluid in
the inner core and an $S$-wave neutron superfluid in the outer
core. In this configuration, each integer vortex in the $S$-wave
superfluid connects to two half-quantized vortices in the $P$-wave
superfluid via topological structures known as
“boojums''~\cite{Marmorini:2020zfp}.  The crust–core interface may act
as a trigger for glitches because it can serve as a barrier for proton
flux tubes dressing neutron vortices, preventing their continuous
annihilation. Furthermore, on secular timescales, the interface
between the $S$-wave and $P$-wave superfluids can radiate energy due
to the time-dependent Josephson current induced by vortex plus
flux-tube motion. Such Josephson current arises because the phases of
$S$-wave and $P$-wave superfluids differ on both side of their
interface~\cite{Sedrakian_Rau2025}.

The microphysics of pairing plays a central role in determining which
regions of a neutron star participate in glitch dynamics. Since
superfluidity (and superconductivity) is a prerequisite for
vortex-mediated angular momentum transfer, any region where pairing is
suppressed is effectively excluded from contributing to glitches. For
example, strong magnetic fields may quench proton superconductivity in
parts of the core, thereby modifying or eliminating flux tube–vortex
interactions and altering the coupling between components. At the same
time, vortex creep models are directly sensitive to the underlying
pairing properties, as the pinning energies and mobility of vortices
depend on the magnitude of the pairing gap.

If the neutron pairing gap is reduced or suppressed in the crust, the
available superfluid reservoir there may be insufficient to account
for observed glitch amplitudes, implying that the core superfluid must
play a partial or even dominant role in glitch activity and
postglitch relaxation. It is important to note that pairing in
neutron stars can, to a good approximation, be described within the
BCS (weak-coupling) regime. A strong-coupling regime—associated with
very low densities and large pairing gaps—is only marginally realized
near the low-density boundary between the inner and outer crust, and
does not play a significant role in the bulk of the star.

\subsection{Long-term variabilities}

Apart from glitches, neutron stars may exhibit complex long-term
rotational variability arising from the interplay between their solid crust and superfluid interior. The classical-like free precession, familiar from rigid-body dynamics, involves the slow rotation of the symmetry axis about the total angular momentum vector. In stars with weakly coupled superfluids, this mode can persist for months or years with the frequency essentially that of classical precession:
\bea \omega_{\rm slow} \simeq \epsilon \Omega,
\eea
$\epsilon$ is the eccentricity, and $\Omega$ is the
stellar rotation rate.  Superimposed on this is a fast precession mode associated with the independent motion of the neutron superfluid. Its frequency scales as
\bea \omega_{\rm fast} \propto \frac{I_s}{I_n} \Omega.
\eea
This mode arises because the superfluid vortices introduce additional degrees of freedom relative to a classical rigid
body~\cite{Sedrakian:1998vi,Link2003,Jones:2016oyh}.

Tkachenko waves provide another source of long-term variability in neutron stars~\cite{Ruderman1970,Noronha:2007qf}. 
The Tkachenko mode frequency in the non-dissipative limit is given in units of $2\Omega$ by
\bea
p_T=\pm \sqrt{\cos\theta+g},
\eea
where $\theta$ is the angle
between the rotation (spin) axis--i.e. the vortex-line direction--and the wave vector $\veck$ and 
$g\equiv({k^2}/{4 \Omega^2 \rho_n})\left[\mu_n-\cos\theta\left(\mu_n-2 \Omega \lambda_V\right)\right]$, where $\rho_n$ is the superfluid density, $\lambda_V$ is the vortex tension, and $\mu_n= \rho_n \hbar \Omega / 8 m_n$ is the shear modulus of the triangular vortex lattice. It is useful to consider the limiting cases of the angular dependence of the Tkachenko modes, specifically in the case of parallel propagation ($\theta=0$):
\bea
  p_T=\pm\left( 1+\frac{k^2}{2\Omega\rho_n}\lambda_V \right)^{1/2},
  \eea
  showing that the mode is dominated by vortex-line tension.
In the case of perpendicular propagation ($\theta=\pi/2$):
  \bea
  p_T=\pm \frac{k}{2\Omega}
  \sqrt{\frac{\mu_n}{\rho_n}},
  \eea
which gives the standard Tkachenko elastic scaling. 

For a general $\theta$ the $\cos\theta$ term acts as a geometric “gap”, while the $k^2$ term encodes lattice elasticity and vortex tension.
Physically, this reflects the crossover from transverse shear
oscillations of the vortex lattice (for
$\veck \perp\boldsymbol{\Omega}$) to longitudinal distortions
along vortex lines (for $\veck\parallel\boldsymbol{\Omega}$),
which is exactly what one expects in a rotating superfluid or
supersolid. 

We note that in a uniformly rotating system, the full low-frequency spectrum contains in addition
to Tkachenko modes (originating from elastic shear deformations of the vortex lattice, with restoring forces provided by the lattice shear modulus and vortex line tension), the inertial modes, which exist without a vortex lattice. They are restored purely by the Coriolis force and have frequencies
$p_I = \pm \cos\theta$ in units of $ 2\Omega$. These modes are non-dispersive, i.e., independent of $k$ at leading order, and vanish for $\veck\perp\boldsymbol{\Omega}$. 

The influence of mutual friction on Tkachenko modes in liquid $^{4}$He
was first studied in~\cite{ChandlerBaym1986}. This framework was later
generalized to include the effects of shear viscosity in the normal
component and applied to neutron stars' interiors
in~\cite{Noronha:2007qf} using Newtonian dissipative superfluid
hydrodynamics. In that context, Tkachenko modes were found to have
characteristic periods ranging from months to years, depending on the
rotation rate $\Omega$, naturally matching the long-term periodicities
observed in pulsars. The observable variability is further shaped by
mutual friction between vortices and the normal fluid, which damps
both precessional and Tkachenko motions.

In the weak-coupling regime, one expects combined signatures of long
postglitch relaxation times together with the possibility of
quasi-free precession. In contrast, in the strong-coupling limit,
precession is strongly suppressed due to efficient coupling between
the superfluid and the normal component, although long-term glitch
relaxation can still persist. Tkachenko modes—low-frequency
oscillations of the vortex lattice—are likewise sensitive to the
dissipative properties of the medium: they are significantly modified
by mutual friction and shear viscosity~\cite{Noronha:2007qf}, while
bulk viscosity does not enter the characteristic equations governing
these modes and therefore has no direct impact on their dynamics.

In our review, we adopted a Newtonian multi-fluid hydrodynamic
framework, which captures the essential features of superfluid
dynamics, vortex motion, and the coupling between superfluid and
normal components relevant for glitch phenomena. This approach
provides a transparent and tractable description of the underlying
physics. However, fully relativistic formulations of superfluid
hydrodynamics do exist~\cite{CarterLanglois1995,Langlois1998}.  More
recently, frameworks~\cite{Gusakov2016,Gusakov2016a,Rau2020} have been
developed to account for finite temperature and/or multiple
components, incorporating vortices, flux tubes, and mutual friction;
such formulations link microscopic physics to macroscopic dynamics,
keeping the general-relativistic framework intact on all scales.

\section{Spin, superfluidity, and vorticity in quark matter}
\label{sec:Quark_Superfluidity}

\subsection{Color superconductivity in quark matter}
{
Building on the discussion of pairing in nucleonic matter, similar
pairing phenomena are expected to occur in deconfined quark matter at
sufficiently high densities, where quarks form Cooper pairs due to
attractive channels in the underlying QCD interaction; for a broad
overview, see Alford {\it et al.}~\cite{Alford2008}. This leads to the emergence of color-superconducting phases, whose structure depends on the number of quark flavors participating, their masses, and the mismatch between their Fermi surfaces imposed by beta equilibrium. The simplest and most robust phase at intermediate densities is the two-flavor color-superconducting (2SC) phase~\cite{Bailin1984}, in which up and down quarks pair in a color-antitriplet channel, leaving one color unpaired. 

At asymptotically high densities, where the strange quark becomes effectively light and Fermi momenta are nearly equal, the CFL phase is expected to be the ground state~\cite{Alford1999}. In this phase, all three flavors participate symmetrically in pairing, leading to a fully gapped spectrum and peculiar properties such as the absence of electrons (in the simplest approximation). Kaon condensation can reintroduce electrons once stress due to finite strange quark mass or charge neutrality drives the system away from exact symmetry, leading to the breaking of the corresponding flavor symmetries and the onset of meson-condensed phases~\cite{Kaplan2002,Bedaque2002}.

At densities relevant for neutron stars, however, these idealized
pairing patterns are further modified by the stress induced by finite
strange quark mass and chemical potential differences between
flavors. This can give rise to phases with broken spatial symmetries,
such as Fulde–Ferrell–type phases where Cooper pairs carry nonzero
momentum~\cite{Anglani2014}, or phases with
deformed Fermi surfaces~\cite{Sedrakian:2003tr}.

These possibilities highlight that, much like in nucleonic matter, the
pairing structure in quark matter is rich and sensitive to microscopic
conditions. The resulting phases can significantly affect transport
and neutrino emission, and therefore play an important role in
determining the observable properties of hybrid
stars~\cite{Alford2008,Grigorian2005}.  }

{Somewhat similarly to nucleonic matter, in color-superconducting
  phases of dense quark matter, magnetic fields couple to quark spins
  and electric charges, leading to Fermi surface mismatches between
  pairing species and thereby suppressing Cooper pairing via a Pauli
  paramagnetic–type mechanism; in this context, the magnetic field and
  the finite strange quark mass play competing and qualitatively
  similar disruptive roles. An additional parallel with nucleonic
  matter arises from the coupling of charged quarks to the magnetic
  field, which induces Landau quantization and modifies the pairing
  structure.  These effects are particularly relevant for phases such
  as 2SC and CFL, where it has been shown that magnetic fields can
  significantly alter the gap structure, induce anisotropies, and even
  lead to the emergence of novel pairing patterns. In particular,
  studies of the CFL phase in a magnetic field demonstrate the
  splitting of pairing gaps and the appearance of magnetically
  modified CFL phase, while analyses of more general quark matter
  phases indicate that magnetic fields can favor pseudoscalar
  condensates and modify the phase diagram in a nontrivial
  way~\cite{Warringa2012,Noronha2007}.  }

{
In color-superconducting quark matter, the response to magnetic fields is more subtle than in conventional superconductors due to the mixing between the electromagnetic gauge field and a component of the gluon field. In particular, the diquark condensate breaks the original $U(1)_{\mathrm{em}}$ symmetry, but leaves unbroken a rotated $U(1)_{\tilde{Q}}$ symmetry corresponding to a linear combination of the photon $A_\mu$ and the 8th gluon $G_\mu^8$. The resulting massless gauge field,
\bea
A_\mu^{\tilde{Q}}=A_\mu \cos \theta-G_\mu^8 \sin \theta,
\eea
defines the so-called rotated photon, while the orthogonal combination,
\bea
A_\mu^X=A_\mu \sin \theta+G_\mu^8 \cos \theta,
\eea
acquires a Meissner mass. The mixing angle $\theta$ is determined by the electromagnetic and strong coupling constants, typically satisfying $\tan \theta \sim e / g \ll 1$, so that the rotated photon is predominantly electromagnetic~\cite{Alford2010}.

As a consequence, magnetic fields associated with the unbroken $U(1)_{\tilde{Q}}$ can penetrate the medium, while those coupled to the massive $X$ field are screened. In the 2SC phase, this leads to partial penetration of magnetic flux and a modified Meissner effect, rather than complete expulsion as in ordinary superconductors. In the color-flavor-locked (CFL) phase, the response to magnetic fields is governed by a similar mixing between the electromagnetic gauge field and a gluon, but with an important difference arising from the fully symmetric participation of all three flavors in pairing. A key distinction of the CFL phase is that all quark quasiparticles are neutral under the rotated charge $\tilde{Q}$. As a result, the medium is an insulator with respect to $U(1)_{\tilde{Q}}$, and no electrons are required to ensure charge neutrality in the idealized limit. Magnetic fields associated with the massless rotated photon can therefore penetrate the CFL phase essentially unimpeded, while the orthogonal $X$ component is screened over a short length scale. Consequently, the CFL phase does not exhibit a conventional Meissner effect for ordinary magnetic fields, but instead allows partial penetration determined by the projection onto the unbroken $U(1)_{\tilde{Q}}$ sector.
}

\subsection{Quantum vorticity in quark matter}

If the color-superconducting phase is of type II with respect to the $X$ field, the corresponding lower critical field is very large, $H_{c 1}^{(X)} \sim 10^{17} \mathrm{G}$~\cite{Alford2010}. Nevertheless, this does not preclude the formation of color-magnetic flux tubes. During the transition into the superconducting phase, $X$-flux can become trapped between growing superconducting domains and subsequently compressed. As these domains merge, the local magnetic field in the remaining normal regions can exceed $H_{c 1}^{(X)}$, leading to their fragmentation into Abrikosov-like flux tubes that carry quantized $X$-flux.

 In contrast, in the CFL phase all quark quasiparticles are neutral under the rotated charge $\tilde{Q}$, so that the medium behaves as an insulator with respect to the unbroken $U(1)_{\tilde{Q}}$, and magnetic fields associated with the rotated photon can penetrate essentially unimpeded, while the orthogonal $X$ component remains screened. Furthermore,  in the CFL phase, breaking of baryon number implies superfluidity, so vortices must appear under rotation; however, the energetically favored defects are not simple $U(1)_B$ vortices, as discussed in the nucleonic case, but non-Abelian {\it semi-superfluid} vortices that carry both fractional circulation and color-magnetic flux~\cite{Balachandran:2005ev}. The CFL matter supports semi-superfluid strings/flux tubes, and later work clarified that the minimal CFL vortices are (1/3)-quantized objects with long-range repulsion, suggesting the formation of a vortex lattice. These vortices also support nontrivial low-energy degrees of freedom in their cores, including orientational modes and, in some treatments, gapless fermionic modes.

 Vorticity has been experimentally observed in the quark–gluon plasma
 created in relativistic heavy-ion collisions~\cite{STAR:2017ckg};
 however, unlike the quantized vortices in superfluids, vorticity in
 the quark–gluon plasma is a property of a relativistic fluid without
 long-range phase coherence, and therefore differs fundamentally in
 its microscopic origin despite some formal analogies.

 An important question for neutron stars is how such quark vortices
 connect to hadronic vortices across a hadron–quark interface. Earlier
 work on colorful boojums showed that junction structures can arise
 when neutron and proton vortices from hadronic matter penetrate a CFL
 core, with the boojum acting as the interface where hadronic
 circulation is redistributed into color-carrying CFL
 vortices\cite{Cipriani:2012hr}. Related studies~\cite{Chatterjee2018}
 discuss vortex continuity and argue that, in crossover scenarios,
 hadronic vortices may connect to non-Abelian CFL
 vortices. Ref.~\cite{Alford2019} showed that in three-flavor
 symmetric matter, a singly quantized hadronic vortex can connect
 smoothly to a singly quantized non-Abelian CFL vortex, so a boojum is
 not always required. Because CFL vortices carry color-magnetic flux,
 they are often described as simultaneously superfluid and magnetic
 defects~\cite{Eto2014}. Altogether, the vortex sector of quark matter
 is likely to play a central role in determining the rotational and
 magnetic properties of hybrid stars. Altogether, the vortex sector of
 quark matter is likely to be central for the rotational, magnetic,
 and transport properties of hybrid stars. In particular, in the 2SC
 phase, color-magnetic flux tubes interact with gapless fermionic
 excitations via the Aharonov–Bohm scattering~\cite{Alford2010},
 leading to a significant drag force that can strongly influence their
 dynamics and mobility, and thereby affect magnetic field evolution
 and transport properties in the stellar core.

In quark matter phases, there is a close analogy to vortex pinning in
neutron-star hadronic phases, where vortices interact with the nuclear
lattice in the crust or with proton flux tubes in the core. It turns
out that in phases with broken spatial symmetries, such as
Fulde–Ferrell or more general crystalline
phases~\cite{Mannarelli2007}, the structure of vortices becomes even
more intricate. In these phases, the pairing gap varies periodically
in space, and the order parameter carries finite
momentum. Consequently, vortices are embedded in an inhomogeneous
background, leading to anisotropic cores and modified tension and
mobility. The periodic modulation of the condensate can act as an
intrinsic pinning lattice for vortices, while the presence of rotated
electromagnetism implies that magnetic flux and superfluid circulation
may be intertwined in a nontrivial way. These features suggest that
quark matter phases with broken spatial symmetries can exhibit
qualitatively new vortex and flux-tube dynamics, with potential
implications for angular momentum transport and glitch phenomena in
hybrid stars.

\section{Conclusions}

We have reviewed selected aspects of the interior physics of compact
stars, with particular emphasis on the microscopic and macroscopic
manifestations of spin, magnetic fields, and nucleonic superfluidity
and superconductivity. Spin plays a fundamental role in the stability
of neutron stars, which are supported by the degeneracy pressure of
fermionic matter. In describing the EoS of dense
matter, we adopted a phenomenological (metamodeling) approach based
on an expansion of the nuclear-matter energy density around the
isospin-symmetric limit and near saturation density. 
This framework
enables the generation of families of EoSs that can be employed in the
general-relativistic stellar-structure equations to determine global
stellar observables such as mass, radius, and moment of inertia.

We reviewed the current state of observational constraints and discussed
prospects for improving them through forthcoming multimessenger
observations. We also examined the influence of magnetic fields on the
EoS of dense matter, noting that significant modifications require
extremely strong fields and may involve the emergence of
spin-polarized or ferromagnetic phases—possibilities that remain
highly uncertain at the densities relevant to neutron-star interiors.

At the mesoscopic scale, the interplay between superfluid and
superconducting components gives rise to a rich variety of vortex and
flux-tube phenomena. Neutron superfluid rotation proceeds through the
formation of a quantized vortex lattice, while protons form a type-II
superconductor characterized by critical fields consistent with those
expected in neutron-star cores. Pinning interactions—between vortices
and nuclei in the crust or between vortices and magnetic flux tubes in
the core—provide a natural mechanism for understanding the
non-stationary rotational dynamics of neutron stars.

Outstanding issues include the internal structure of
${}^3P_2$–${}^3F_2$ vortices, the possible emergence of type-I
superconductivity in the deep core, the configuration and dynamics of
proton flux tubes, the magnitude of mutual friction coefficients in
different pairing regimes, and the extent of crustal entrainment. The
dynamics at the mesoscopic scale connects directly to macroscopic
models of pulsar glitches and postglitch relaxation, which remain
unresolved regarding the precise location and extent of the superfluid
regions responsible for short- and long-term relaxation
processes. While several generic glitch triggers are qualitatively
understood, a quantitative description remains incomplete. Major open
problems include identifying the location and nature of glitch
activity (crustal, core, or mixed), determining the components
responsible for postglitch recovery, and explaining the observed
diversity of glitch behavior.

We have also reviewed the possible role of deconfined quark matter and
color superconductivity in the inner cores of hybrid stars. The rich
pairing structure of dense QCD gives rise to a variety of
superconducting phases, each with distinct transport and magnetic
properties. In particular, color-superconducting phases support novel
vortex configurations, including color-magnetic flux tubes in the 2SC
phase and non-Abelian semi-superfluid vortices in the CFL phase, which
intertwine superfluid circulation with rotated electromagnetic
flux. In the CFL-like phases, these structures may play an important
role in the rotational dynamics of hybrid stars, providing additional
channels for angular momentum transfer and mutual friction, through
such mechanisms as scattering of electrons and quarks off
color-magnetic defects via Aharonov–Bohm interactions.  At the same
time, the modified electromagnetic response of color-superconducting
matter, governed by rotated gauge fields, may affect the evolution and
topology of magnetic fields in the stellar core. While these
possibilities remain model-dependent and subject to significant
uncertainties in the QCD phase diagram at neutron-star densities, they
underscore the potential of quark matter to leave observable imprints
on both the rotational and magnetic behavior of compact stars.

In conclusion, ongoing and forthcoming multimessenger
observations—across the electromagnetic spectrum and through
gravitational-wave detections—are expected to provide crucial insights
into the structure and dynamics of neutron stars. These observations
will offer essential feedback for refining theoretical models, linking
microscopic nuclear interactions to macroscopic astrophysical
observables, and thereby deepening our understanding of strongly
interacting matter under extreme conditions.


\clearpage 

%
\bibliographystyle{sciencemag}

\begin{thebibliography}{300}
\providecommand{\url}[1]{\texttt{#1}}
\expandafter\ifx\csname urlstyle\endcsname\relax
  \providecommand{\doi}[1]{doi:\discretionary{}{}{}#1}\else
  \providecommand{\doi}{doi:\discretionary{}{}{}\begingroup
    \urlstyle{rm}\Url}\fi
  

\bibitem{UhlenbeckGoudsmit1926}
G.~E. Uhlenbeck, S.~A. Goudsmit, Spinning Electrons and the Structure of
  Spectra. \emph{Nature} \textbf{117}, 264--265 (1926), english announcement of
  the spin hypothesis (20 Feb 1926)., \doi{10.1038/117264a0}.

\bibitem{Pauli1925}
W.~Pauli, Ueber den Zusammenhang des Abschlusses der Elektronengruppen im Atom
  mit der Komplexstruktur der Spektren. \emph{Zeitschrift f\"ur Physik}
  \textbf{31}, 765--783 (1925), pauli's original paper introducing the
  exclusion principle., \doi{10.1007/BF02980631}.

\bibitem{Pauli1927}
W.~Pauli, Zur Quantenmechanik des magnetischen Elektrons. \emph{Zeitschrift
  f\"ur Physik} \textbf{43}, 601--623 (1927), introduces spinor formalism and
  the Pauli spin matrices., \doi{10.1007/BF01397326}.

\bibitem{Dirac1928}
P.~A.~M. Dirac, The Quantum Theory of the Electron. \emph{Proceedings of the
  Royal Society of London. Series A, Containing Papers of a Mathematical and
  Physical Character} \textbf{117}~(778), 610--624 (1928), dirac's original
  paper introducing the relativistic wave equation for the electron.,
  \doi{10.1098/rspa.1928.0023}.

\bibitem{Lattimer2004Sci}
J.~M. {Lattimer}, M.~{Prakash}, {The Physics of Neutron Stars}. \emph{Science}
  \textbf{304}, 536--542 (2004), \doi{10.1126/science.1090720}.

\bibitem{Sedrakian2023PrPNP}
A.~{Sedrakian}, J.~J. {Li}, F.~{Weber}, {Heavy baryons in compact stars}.
  \emph{Prog. Part. Nucl. Phys.} \textbf{131}, 104041 (2023),
  \doi{10.1016/j.ppnp.2023.104041}.

\bibitem{Tolman1939}
R.~C. Tolman, Static Solutions of Einstein's Field Equations for Spheres of
  Fluid. \emph{Phys. Rev.} \textbf{55}, 364--373 (1939).

\bibitem{Oppenheimer1939}
J.~R. Oppenheimer, G.~M. Volkoff, On Massive Neutron Cores. \emph{Phys. Rev.}
  \textbf{55}, 374--381 (1939).

\bibitem{Landau1980}
L.~D. {Landau}, E.~M. {Lifshitz}, \emph{Statistical physics, Part 1} (Pergamon
  Press, New York) (1980).

\bibitem{Shapiro1964}
I.~I. Shapiro, Fourth Test of General Relativity. \emph{Phys. Rev. Lett.}
  \textbf{13}, 789--791 (1964), \doi{10.1103/PhysRevLett.13.789},
  \url{https://link.aps.org/doi/10.1103/PhysRevLett.13.789}.

\bibitem{Demorest:2010bx}
P.~Demorest, T.~Pennucci, S.~Ransom, M.~Roberts, J.~Hessels, {Shapiro Delay
  Measurement of A Two Solar Mass Neutron Star}. \emph{Nature} \textbf{467},
  1081--1083 (2010), \doi{10.1038/nature09466}.

\bibitem{Arzoumanian2018}
Z.~Arzoumanian, \emph{et~al.}, {The NANOGrav 11-year Data Set: High-precision
  timing of 45 Millisecond Pulsars}. \emph{Astrophys. J. Suppl.}
  \textbf{235}~(2), 37 (2018), \doi{10.3847/1538-4365/aab5b0}.

\bibitem{Cromartie:2019kug}
H.~T. Cromartie, \emph{et~al.}, {Relativistic Shapiro delay measurements of an
  extremely massive millisecond pulsar}. \emph{Nature Astron.} \textbf{4}~(1),
  72--76 (2020).

\bibitem{Antoniadis:2013pzd}
J.~Antoniadis, \emph{et~al.}, {A Massive Pulsar in a Compact Relativistic
  Binary}. \emph{Science} \textbf{340}, 6131 (2013).

\bibitem{Fonseca:2021wxt}
E.~Fonseca, \emph{et~al.}, {Refined Mass and Geometric Measurements of the
  High-mass PSR J0740+6620}. \emph{Astrophys. J. Lett.} \textbf{915}~(1), L12
  (2021), \doi{10.3847/2041-8213/ac03b8}.

\bibitem{Bardeen1966}
J.~M. {Bardeen}, K.~S. {Thorne}, D.~W. {Meltzer}, {A Catalogue of Methods for
  Studying the Normal Modes of Radial Pulsation of General-Relativistic Stellar
  Models}. \emph{Astrophys. J.} \textbf{145}, 505 (1966), \doi{10.1086/148791}.

\bibitem{AbbottPhysRevX.9.011001}
B.~P. Abbott, \emph{et~al.}, Properties of the Binary Neutron Star Merger
  GW170817. \emph{Phys. Rev. X} \textbf{9}, 011001 (2019),
  \doi{10.1103/PhysRevX.9.011001},
  \url{https://link.aps.org/doi/10.1103/PhysRevX.9.011001}.

\bibitem{Abbott2020}
B.~P. {Abbott}, \emph{et~al.}, {GW190425: Observation of a Compact Binary
  Coalescence with Total Mass {\ensuremath{\sim}} 3.4
  M$_{{\ensuremath{\odot}}}$}. \emph{Astrophys. J. Lett.} \textbf{892}~(1), L3
  (2020), \doi{10.3847/2041-8213/ab75f5}.

\bibitem{flanagan2008constraining}
E.~E. Flanagan, T.~Hinderer, Constraining neutron-star tidal Love numbers with
  gravitational-wave detectors. \emph{Phys. Rev. D—Particles, Fields,
  Gravitation, and Cosmology} \textbf{77}~(2), 021502 (2008).

\bibitem{Hinderer:2007}
T.~Hinderer, {Tidal Love numbers of neutron stars}. \emph{Astrophys. J.}
  \textbf{677}, 1216--1220 (2008), \doi{10.1086/533487}.

\bibitem{NICER:2019a}
T.~E. Riley, A.~L. Watts, S.~Bogdanov, \emph{et~al.}, {A NICER View of PSR
  J0030+0451: Millisecond Pulsar Parameter Estimation}. \emph{Astrophy. J.
  Lett.} \textbf{887}~(1), L21 (2019), \doi{10.3847/2041-8213/ab481c}.

\bibitem{NICER:2019b}
M.~C. Miller, F.~K. Lamb, A.~J. Dittmann, \emph{et~al.}, {PSR J0030+0451 Mass
  and Radius from NICER Data and Implications for the Properties of Neutron
  Star Matter}. \emph{Astrophys. J. Lett.} \textbf{887}~(1), L24 (2019),
  \doi{10.3847/2041-8213/ab50c5}.

\bibitem{NICER:2021a}
T.~E. Riley, A.~L. Watts, P.~S. Ray, \emph{et~al.}, {A NICER View of the
  Massive Pulsar PSR J0740+6620 Informed by Radio Timing and XMM-Newton\
  Spectroscopy}. \emph{Astrophy. J. Lett.} \textbf{918}~(2), L27 (2021),
  \doi{10.3847/2041-8213/ac0a81}.

\bibitem{NICER:2021b}
M.~C. Miller, F.~K. Lamb, A.~J. Dittmann, \emph{et~al.}, {The Radius of PSR
  J0740+6620 from NICER and XMM-Newton Data}. \emph{Astrophy. J. Lett.}
  \textbf{918}~(2), L28 (2021), \doi{10.3847/2041-8213/ac089b}.

\bibitem{Vinciguerra2024}
S.~Vinciguerra, \emph{et~al.}, An Updated Mass–Radius Analysis of the
  2017–2018 NICER Data Set of PSR J0030+0451. \emph{Astrophys. J.}
  \textbf{961}, 62 (2024), published 16 January 2024,
  \doi{10.3847/1538-4357/acfb83}.

\bibitem{Kramer2021}
M.~Kramer, \emph{et~al.}, Strong-Field Gravity Tests with the Double Pulsar.
  \emph{Phys. Rev. X} \textbf{11}~(4), 041050 (2021),
  \doi{10.1103/PhysRevX.11.041050}.

\bibitem{LattimerSchutz2004}
J.~M. Lattimer, B.~F. Schutz, Constraining the Equation of State with Moment of
  Inertia Measurements. \emph{Astrophys. J.} \textbf{629}, 979--984 (2005),
  \doi{10.1086/431543}.

\bibitem{Gandolfi2012}
S.~Gandolfi, J.~Carlson, S.~Reddy, The maximum mass and radius of neutron stars
  and the nuclear symmetry energy. \emph{Phys. Rev. C} \textbf{85}~(3), 032801
  (2012), \doi{10.1103/PhysRevC.85.032801}.

\bibitem{Yang:2025}
Y.~Yang, N.~C. Camacho, M.~Hippert, J.~Noronha-Hostler, Symmetry-energy
  expansion with strange dense matter. \emph{Phys. Rev. C} \textbf{113}, 045805 (2026),
  \doi{10.1103/trk9-8gph}, \url{https://link.aps.org/doi/10.1103/trk9-8gph}.

\bibitem{Imam2022}
S.~M.~A. Imam, N.~K. Patra, C.~Mondal, T.~Malik, B.~K. Agrawal, Bayesian
  reconstruction of nuclear matter parameters from the equation of state of
  neutron star matter. \emph{Phys. Rev. C} \textbf{105}, 015806 (2022),
  \doi{10.1103/PhysRevC.105.015806},
  \url{https://link.aps.org/doi/10.1103/PhysRevC.105.015806}.

\bibitem{Xie:2020kta}
W.-J. Xie, B.-A. Li, {Bayesian inference of the incompressibility, skewness and
  kurtosis of nuclear matter from empirical pressures in relativistic heavy-ion
  collisions}. \emph{J. Phys. G: Nucl. Part. Phys.} \textbf{48}~(2), 025110
  (2021), \doi{10.1088/1361-6471/abd25a}.

\bibitem{Li:2025oxi}
J.-J. Li, Y.~Tian, A.~Sedrakian, {Bayesian inferences on covariant density
  functionals from multi-messenger astrophysical data: Nucleonic models}.
  \emph{Phys. Rev. C} \textbf{111}~(5), 055804 (2025),
  \doi{10.1103/PhysRevC.111.055804}.

\bibitem{Li:2025vhk}
J.-J. Li, A.~Sedrakian, {Bayesian inferences on covariant density functionals
  from multi-messenger astrophysical data: The impacts of likelihood functions
  of low density matter constraints}. \emph{Phys. Rev. C} \textbf{112}~(1),
  015802 (2025), \doi{10.1103/c1k3-k4l5}.

\bibitem{Margueron:2018eob}
J.~Margueron, F.~Gulminelli, {Effect of high-order empirical parameters on the
  nuclear equation of state}. \emph{Phys. Rev. C} \textbf{99}~(2), 025806
  (2019), \doi{10.1103/PhysRevC.99.025806}.

\bibitem{Marczenko:2020jma}
M.~Marczenko, D.~Blaschke, K.~Redlich, C.~Sasaki, {Toward a unified equation of
  state for multi-messenger astronomy}. \emph{Astron. Astrophys.} \textbf{643},
  A82 (2020), \doi{10.1051/0004-6361/202038211}.

\bibitem{Koehn2025}
H.~Koehn, \emph{et~al.}, From Existing and New Nuclear and Astrophysical
  Constraints to Stringent Limits on the Equation of State of Neutron-Rich
  Dense Matter. \emph{Phys. Rev. X} \textbf{15}, 021014 (2025),
  \doi{10.1103/PhysRevX.15.021014},
  \url{https://link.aps.org/doi/10.1103/PhysRevX.15.021014}.

\bibitem{Li:2023bid}
J.~J. Li, A.~Sedrakian, {New Covariant Density Functionals of Nuclear Matter
  for Compact Star Simulations}. \emph{Astrophys. J.} \textbf{957}~(1), 41
  (2023), \doi{10.3847/1538-4357/acfa73}.

\bibitem{LIGO-Virgo2020}
R.~Abbott, T.~D. Abbott, S.~Abraham, \emph{et~al.}, {GW190814: Gravitational
  Waves from the Coalescence of a 23 Solar Mass Black Hole with a 2.6 Solar
  Mass Compact Object}. \emph{Astrophys. J. Lett.} \textbf{896}~(2), L44
  (2020), \doi{10.3847/2041-8213/ab960f}.

\bibitem{Bombaci2017}
I.~{Bombaci}, {The Hyperon Puzzle in Neutron Stars}, J. Phys. Soc. Jpn. Conf. Proc. 17, 101002 (2017), \doi{10.7566/JPSCP.17.101002}.

\bibitem{Oertel2017}
M.~{Oertel}, M.~{Hempel}, T.~{Kl{\"a}hn}, S.~{Typel}, {Equations of state for
  supernovae and compact stars}. \emph{Rev. Mod. Phys.} \textbf{89}~(1), 015007
  (2017), \doi{10.1103/RevModPhys.89.015007}.

\bibitem{Prakash1992}
M.~{Prakash}, M.~{Prakash}, J.~M. {Lattimer}, C.~J. {Pethick}, {Rapid Cooling
  of Neutron Stars by Hyperons and Delta Isobars}. \emph{Astrophys. J. Lett.}
  \textbf{390}, L77 (1992), \doi{10.1086/186376}.

\bibitem{Fortin2021}
M.~{Fortin}, A.~R. {Raduta}, S.~{Avancini}, C.~{Provid{\^e}ncia}, {Thermal
  evolution of relativistic hyperonic compact stars with calibrated equations
  of state}. \emph{Phys. Rev. D} \textbf{103}~(8), 083004 (2021),
  \doi{10.1103/PhysRevD.103.083004}.

\bibitem{Raduta2019}
A.~R. {Raduta}, J.~J. {Li}, A.~{Sedrakian}, F.~{Weber}, {Cooling of
  hypernuclear compact stars: Hartree-Fock models and high-density pairing}.
  \emph{MNRAS} \textbf{487}~(2), 2639--2652 (2019),
  \doi{10.1093/mnras/stz1459}.

\bibitem{Providencia2019FrASS}
C.~{Provid{\^e}ncia}, M.~{Fortin}, H.~{Pais}, A.~{Rabhi}, {Hyperonic stars and
  the symmetry energy}. \emph{Front. Astron. Space Sci.} \textbf{6}, 13 (2019),
  \doi{10.3389/fspas.2019.00013}.

\bibitem{Alford2013}
M.~G. {Alford}, S.~{Han}, M.~{Prakash}, {Generic conditions for stable hybrid
  stars}. \emph{Phys. Rev. D} \textbf{88}~(8), 083013 (2013),
  \doi{10.1103/PhysRevD.88.083013}.

\bibitem{Christian2018}
J.-E. {Christian}, A.~{Zacchi}, J.~{Schaffner-Bielich}, {Classifications of
  twin star solutions for a constant speed of sound parameterized equation of
  state}. \emph{Eur. Phys. J. A} \textbf{54}~(2), 28 (2018),
  \doi{10.1140/epja/i2018-12472-y}.

\bibitem{Li2025JCAP}
J.~J. {Li}, A.~{Sedrakian}, M.~{Alford}, {Confronting new NICER mass-radius
  measurements with phase transition in dense matter and twin compact stars}.
  \emph{JCAP} \textbf{2025}~(2), 002 (2025),
  \doi{10.1088/1475-7516/2025/02/002}.

\bibitem{Alford2017}
M.~{Alford}, A.~{Sedrakian}, {Compact Stars with Sequential QCD Phase
  Transitions}. \emph{Phys. Rev. Lett.} \textbf{119}~(16), 161104 (2017),
  \doi{10.1103/PhysRevLett.119.161104}.

\bibitem{Tan2020}
H.~Tan, J.~Noronha-Hostler, N.~Yunes, Neutron Star Equation of State in light
  of GW190814. \emph{Phys. Rev. Lett.} \textbf{125}~(26), 261104 (2020),
  \doi{10.1103/PhysRevLett.125.261104}.

\bibitem{Turolla2015}
R.~{Turolla}, S.~{Zane}, A.~L. {Watts}, {Magnetars: the physics behind
  observations. A review}. \emph{Rep. Prog. Phys.} \textbf{78}~(11), 116901
  (2015), \doi{10.1088/0034-4885/78/11/116901}.

\bibitem{Adhikari2026PrPNP}
P.~{Adhikari}, \emph{et~al.}, {Strongly interacting matter in extreme magnetic
  fields}. \emph{Prog. Part. Nucl. Phys.} \textbf{146}, 104199 (2026),
  \doi{10.1016/j.ppnp.2025.104199}.

\bibitem{Bocquet1995}
M.~{Bocquet}, S.~{Bonazzola}, E.~{Gourgoulhon}, J.~{Novak}, {Rotating neutron
  star models with a magnetic field.} \emph{Astron. Astrophys.} \textbf{301},
  757 (1995).

\bibitem{Broderick2000}
A.~{Broderick}, M.~{Prakash}, J.~M. {Lattimer}, {The Equation of State of
  Neutron Star Matter in Strong Magnetic Fields}. \emph{Astrophys. J.}
  \textbf{537}~(1), 351--367 (2000), \doi{10.1086/309010}.

\bibitem{Cardall2001}
C.~Y. {Cardall}, M.~{Prakash}, J.~M. {Lattimer}, {Effects of Strong Magnetic
  Fields on Neutron Star Structure}. \emph{Astrophys. J.} \textbf{554},
  322--339 (2001), \doi{10.1086/321370}.

\bibitem{Thapa:2021kfo}
V.~B. Thapa, M.~Sinha, J.~J. Li, A.~Sedrakian, {Massive $\Delta$-resonance
  admixed hypernuclear stars with antikaon condensations}. \emph{Phys. Rev. D}
  \textbf{103}~(6), 063004 (2021), \doi{10.1103/PhysRevD.103.063004}.

\bibitem{Peterson:2023bmr}
J.~Peterson, \emph{et~al.}, {Temperature and strong magnetic field effects in
  dense matter}. \emph{Phys. Rev. D} \textbf{108}~(6), 063011 (2023),
  \doi{10.1103/PhysRevD.108.063011}.

\bibitem{Most2025ApJ}
E.~R. {Most}, J.~{Peterson}, L.~{Scurto}, H.~{Pais}, V.~{Dexheimer}, {Impact of
  Magnetic-field-driven Anisotropies on the Equation of State Probed in Neutron
  Star Mergers}. \emph{Astrophys. J. Lett.} \textbf{989}~(2), L29 (2025),
  \doi{10.3847/2041-8213/adf62d}.

\bibitem{Lai2001}
D.~{Lai}, {Matter in strong magnetic fields}. \emph{Rev. Mod. Phys.}
  \textbf{73}~(3), 629–662 (2001), \doi{10.1103/RevModPhys.73.629}.

\bibitem{Harding2006}
A.~K. Harding, D.~Lai, Physics of Strongly Magnetized Neutron Stars. \emph{Rep.
  Prog. Phys.} \textbf{69}~(9), 2631--2708 (2006),
  \doi{10.1088/0034-4885/69/9/R03}.

\bibitem{Chatterjee2021Review}
D.~Chatterjee, J.~Novak, M.~Oertel, {Structure of ultra-magnetised neutron
  stars}. \emph{Eur. Phys. J. A} \textbf{57}~(8), 249 (2021),
  \doi{10.1140/epja/s10050-021-00525-5}.

\bibitem{Pili2014}
A.~G. {Pili}, N.~{Bucciantini}, L.~{Del Zanna}, {Axisymmetric equilibrium
  models for magnetized neutron stars in General Relativity under the
  Conformally Flat Condition}. \emph{MNRAS} \textbf{439}~(4), 3541--3563
  (2014), \doi{10.1093/mnras/stu215}.

\bibitem{Das:2025fws}
M.~Das, A.~Sedrakian, B.~Mukhopadhyay, {Superconductivity in magnetars:
  Exploring type-I and type-II states in toroidal magnetic fields}. \emph{Phys.
  Rev. D} \textbf{111}~(8), L081307 (2025), \doi{10.1103/PhysRevD.111.L081307}.

\bibitem{Clark1969}
J.~W. Clark, Magnetic Susceptibility of Neutron Matter. \emph{Phys. Rev. Lett.}
  \textbf{23}, 1463--1466 (1969), \doi{10.1103/PhysRevLett.23.1463},
  \url{https://link.aps.org/doi/10.1103/PhysRevLett.23.1463}.

\bibitem{Tews2020ApJ}
I.~{Tews}, A.~{Schwenk}, {Spin-polarized Neutron Matter, the Maximum Mass of
  Neutron Stars, and GW170817}. \emph{Astrophys. J.} \textbf{892}~(1), 14
  (2020), \doi{10.3847/1538-4357/ab7232}.

\bibitem{Vidana2002}
I.~{Vida{\~n}a}, I.~{Bombaci}, {Equation of state and magnetic susceptibility
  of spin polarized isospin asymmetric nuclear matter}. \emph{Phys. Rev. C}
  \textbf{66}~(4), 045801 (2002), \doi{10.1103/PhysRevC.66.045801}.

\bibitem{Haensel1996}
P.~{Haensel}, S.~{Bonazzola}, {Ferromagnetism of dense matter and magnetic
  properties of neutron stars.} \emph{Astron. Astrophys.} \textbf{314},
  1017--1023 (1996), \doi{10.48550/arXiv.astro-ph/9605149}.

\bibitem{Arras:1998mv}
P.~Arras, D.~Lai, {Neutrino - nucleon interactions in magnetized neutron star
  matter: The Effects of parity violation}. \emph{Phys. Rev. D} \textbf{60},
  043001 (1999), \doi{10.1103/PhysRevD.60.043001}.

\bibitem{Maruyama2022}
T.~{Maruyama}, \emph{et~al.}, {A relativistic quantum approach to neutrino and
  antineutrino emission via the direct Urca process in strongly magnetized
  neutron-star matter}. \emph{Phys. Lett. B} \textbf{824}, 136813 (2022),
  \doi{10.1016/j.physletb.2021.136813}.

\bibitem{Huang2010}
X.-G. {Huang}, M.~{Huang}, D.~H. {Rischke}, A.~{Sedrakian}, {Anisotropic
  hydrodynamics, bulk viscosities, and r-modes of strange quark stars with
  strong magnetic fields}. \emph{Phys. Rev. D} \textbf{81}~(4), 045015 (2010),
  \doi{10.1103/PhysRevD.81.045015}.

\bibitem{Potekhin:2015qsa}
A.~Y. Potekhin, J.~A. Pons, D.~Page, {Neutron stars - cooling and transport}.
  \emph{Space Sci. Rev.} \textbf{191}~(1-4), 239--291 (2015),
  \doi{10.1007/s11214-015-0180-9}.

\bibitem{Shovkovy:2025yvn}
I.~A. Shovkovy, R.~Ghosh, {Review of heat and charge transport in strongly
  magnetized relativistic plasmas.}  AAPPS Bull. 35, 34 (2025).

\bibitem{Ventura2001}
J.~{Ventura}, A.~{Potekhin}, {Neutron Star Envelopes and Thermal Radiation from
  the Magnetic Surface}, in \emph{The Neutron Star - Black Hole Connection},
  C.~{Kouveliotou}, J.~{Ventura}, E.~{van den Heuvel}, Eds., vol. 567 (Springer, 2001), 
  p. 393, \doi{10.48550/arXiv.astro-ph/0104003}.

\bibitem{Chamel2015}
N.~{Chamel}, \emph{et~al.}, {Role of Landau quantization on the neutron-drip
  transition in magnetar crusts}. \emph{Phys. Rev. C} \textbf{91}~(6), 065801
  (2015), \doi{10.1103/PhysRevC.91.065801}.

\bibitem{Blandford1982}
R.~D. Blandford, L.~Hernquist, Magnetic Susceptibility of a Neutron Star Crust.
  \emph{J. Phys. C: Solid State Phys.} \textbf{15}~(30), 6233--6243 (1982),
  \doi{10.1088/0022-3719/15/30/017}.

\bibitem{Suh2010}
I.-S. Suh, G.~J. Mathews, Magnetic Domains in Magnetar Matter as an Engine for
  Soft Gamma-Ray Repeaters and Anomalous {{X-ray}} Pulsars. \emph{Astrophys.
  J.} \textbf{717}~(2), 843--848 (2010), \doi{10.1088/0004-637X/717/2/843}.

\bibitem{Rau2023}
P.~B. Rau, I.~Wasserman, Magnetohydrodynamic Stability of Magnetars in the
  Ultrastrong Field Regime {{II}}: {{The}} Crust. \emph{MNRAS} \textbf{520},
  1173--1192 (2023), \doi{10.1093/mnras/stad146}.

\bibitem{Rau2025}
P.~B. {Rau}, I.~{Wasserman}, {Numerical Simulation of Electron
  Magnetohydrodynamics with Landau-quantized Electrons in Magnetar Crusts}.
  \emph{Astrophys. J.} \textbf{979}~(2), 154 (2025),
  \doi{10.3847/1538-4357/ad9dea}.

\bibitem{Sedrakian2019}
A.~{Sedrakian}, J.~W. {Clark}, {Superfluidity in nuclear systems and neutron
  stars}. \emph{Eur. Phys. J. A} \textbf{55}~(9), 167 (2019),
  \doi{10.1140/epja/i2019-12863-6}.

\bibitem{Zverev2003}
M.~V. {Zverev}, J.~W. {Clark}, V.~A. {Khodel}, {$^{3}P_{2}$-$^{3}F_{2}$ pairing
  in dense neutron matter: the spectrum of solutions}. \emph{Nucl. Phys. A}
  \textbf{720}, 20--42 (2003), \doi{10.1016/S0375-9474(03)00653-5}.

\bibitem{ZuoCuiLombardoSchulze2008}
W.~{Zuo}, C.~X. {Cui}, U.~{Lombardo}, H.-J. {Schulze}, {Three-body force effect
  on P3 F$_{2}$ neutron superfluidity in neutron matter, neutron star matter,
  and neutron stars}. \emph{Phys. Rev. C} \textbf{78}~(1), 015805 (2008),
  \doi{10.1103/PhysRevC.78.015805}.

\bibitem{DongLombardZuo2013}
J.~M. {Dong}, U.~{Lombardo}, W.~{Zuo}, {$^{3}$PF$_{2}$ pairing in high-density
  neutron matter}. \emph{Phys. Rev. C} \textbf{87}~(6), 062801 (2013),
  \doi{10.1103/PhysRevC.87.062801}.

\bibitem{Ding2016}
D.~{Ding}, \emph{et~al.}, {Pairing in high-density neutron matter including
  short- and long-range correlations}. \emph{Phys. Rev. C} \textbf{94}~(2),
  025802 (2016), \doi{10.1103/PhysRevC.94.025802}.

\bibitem{Papakonstantinou:2017ewy}
P.~Papakonstantinou, J.~W. Clark, {Three-Nucleon Forces and Triplet Pairing in
  Neutron Matter}. \emph{J. Low Temp. Phys.} \textbf{189}~(5-6), 361--382
  (2017), \doi{10.1007/s10909-017-1808-9}.

\bibitem{AlmSedrakian1996}
T.~{Alm}, G.~{R{\"o}pke}, A.~{Sedrakian}, F.~{Weber}, {$^{3}D_{2}$ pairing in
  asymmetric nuclear matter}. \emph{Nucl. Phys. A} \textbf{604}, 491--504
  (1996), \doi{10.1016/0375-9474(96)00153-4}.

\bibitem{Stein2016}
M.~{Stein}, A.~{Sedrakian}, X.-G. {Huang}, J.~W. {Clark}, {Spin-polarized
  neutron matter: Critical unpairing and BCS-BEC precursor}. \emph{Phys. Rev.
  C} \textbf{93}~(1), 015802 (2016), \doi{10.1103/PhysRevC.93.015802}.

\bibitem{SinhaSedrakian2015}
M.~{Sinha}, A.~{Sedrakian}, {Magnetar superconductivity versus magnetism:
  Neutrino cooling processes}. \emph{Phys. Rev. C} \textbf{91}~(3), 035805
  (2015), \doi{10.1103/PhysRevC.91.035805}.

\bibitem{Haber2017}
A.~{Haber}, A.~{Schmitt}, {Critical magnetic fields in a superconductor coupled
  to a superfluid}. \emph{Phys. Rev. D} \textbf{95}~(11), 116016 (2017),
  \doi{10.1103/PhysRevD.95.116016}.

\bibitem{TinkhamBook}
M.~{Tinkham}, \emph{Introduction to superconductivity,} (McGraw-Hill, 1996).

\bibitem{Muzikar1980}
P.~{Muzikar}, J.~A. {Sauls}, J.~W. {Serene}, {$^{3}$P$_{2}$ pairing in
  neutron-star matter: Magnetic field effects and vortices}. \emph{Phys. Rev.
  D} \textbf{21}~(6), 1494--1502 (1980), \doi{10.1103/PhysRevD.21.1494}.

\bibitem{Sauls1981}
J.~A. {Sauls}, D.~L. {Stein}, {p-wave superfluidity in neutron stars and
  $^{3}$He.} \emph{Physica B: Condensed Matter} \textbf{107}, 55--56 (1981).

\bibitem{Sauls1982}
J.~A. {Sauls}, D.~L. {Stein}, J.~W. {Serene}, {Magnetic vortices in a rotating
  $^{3}$P$_{2}$ neutron superfluid}. \emph{Phys. Rev. D} \textbf{25}~(4),
  967--975 (1982), \doi{10.1103/PhysRevD.25.967}.

\bibitem{Sauls2019}
J.~A. {Sauls}, {Superfluidity in the Interiors of Neutron Stars}. \emph{arXiv
  e-prints} arXiv:1906.09641 (2019), \doi{10.48550/arXiv.1906.09641}.

\bibitem{PitaevskiiBEC}
L.~Pitaevskii, S.~Stringari, \emph{Bose-Einstein Condensation} (Oxford
  University Press) (2003).

\bibitem{Gennes1999superconductivity}
P.-G. de~Gennes, \emph{Superconductivity of Metals and Alloys} (Advanced Book
  Program, Perseus Books, New York, N.Y.) (1999).

\bibitem{Masaki:2021hmk}
Y.~Masaki, T.~Mizushima, M.~Nitta, {Non-Abelian half-quantum vortices in 3P2
  topological superfluids}. \emph{Phys. Rev. B} \textbf{105}~(22), L220503
  (2022), \doi{10.1103/PhysRevB.105.L220503}.

\bibitem{Bedaque:2012bs}
P.~F. Bedaque, A.~N. Nicholson, {Low lying modes of triplet-condensed neutron
  matter and their effective theory}. \emph{Phys. Rev. C} \textbf{87}~(5),
  055807 (2013), [Erratum: Phys.Rev.C 89, 029902 (2014)],
  \doi{10.1103/PhysRevC.87.055807}.

\bibitem{Bedaque:2013fja}
P.~F. Bedaque, S.~Reddy, {Goldstone modes in the neutron star core}.
  \emph{Phys. Lett. B} \textbf{735}, 340--343 (2014),
  \doi{10.1016/j.physletb.2014.06.033}.

\bibitem{Leinson:2010pk}
L.~B. Leinson, {Superfluid phases of triplet pairing and neutrino emission from
  neutron stars}. \emph{Phys. Rev. C} \textbf{82}, 065503 (2010), [Erratum:
  Phys.Rev.C 84, 049901 (2011)], \doi{10.1103/PhysRevC.82.065503}.

\bibitem{Kobayashi:2022moc}
M.~Kobayashi, M.~Nitta, {Core structures of vortices in Ginzburg-Landau theory
  for neutron $^3P_2$ superfluids}. \emph{Phys. Rev. C} \textbf{105}~(3),
  035807 (2022), \doi{10.1103/PhysRevC.105.035807}.

\bibitem{Masaki:2023rtn}
Y.~Masaki, T.~Mizushima, M.~Nitta, {Non-Abelian Anyons and Non-Abelian Vortices
  in Topological Superconductors}, in \emph{{Encyclopedia of Condensed Matter
  Physics (Second Edition)}} ({Elsevier}), vol.~2 (2024), pp. 755--794,
  \doi{10.1016/B978-0-323-90800-9.00225-0}.

\bibitem{Yasui:2019pgb}
S.~Yasui, C.~Chatterjee, M.~Nitta, {Topological defects at the boundary of
  neutron $^{3}P_{2}$ superfluids in neutron stars}. \emph{Phys. Rev. C}
  \textbf{101}~(2), 025204 (2020), \doi{10.1103/PhysRevC.101.025204}.

\bibitem{Sedrakian_Rau2025}
A.~{Sedrakian}, P.~B. {Rau}, {Josephson currents in neutron stars}. \emph{Phys.
  Rev. D} \textbf{111}~(2), 023044 (2025), \doi{10.1103/PhysRevD.111.023044}.

\bibitem{Sedrakian1980}
D.~M. {Sedrakian}, K.~M. {Shakhabasian}, {On a mechanism of magnetic field
  generation in pulsars}. \emph{Astrofizika} \textbf{16}, 727--736 (1980).

\bibitem{Alpar1984ApJ}
M.~A. {Alpar}, S.~A. {Langer}, J.~A. {Sauls}, {Rapid postglitch spin-up of the
  superfluid core in pulsars}. \emph{Astrophys. J.} \textbf{282}, 533--541
  (1984), \doi{10.1086/162232}.

\bibitem{Haskell:2017lkl}
B.~Haskell, A.~Sedrakian, {Superfluidity and Superconductivity in Neutron
  Stars}. \emph{Astrophys. Space Sci. Libr.} \textbf{457}, 401--454 (2018),
  \doi{10.1007/978-3-319-97616-7_8}.

\bibitem{Antonopoulou2022}
D.~{Antonopoulou}, B.~{Haskell}, C.~M. {Espinoza}, {Pulsar glitches:
  observations and physical interpretation}. \emph{Rep. Prog. Phys.}
  \textbf{85}~(12), 126901 (2022), \doi{10.1088/1361-6633/ac9ced}.

\bibitem{Bildsten1989}
L.~{Bildsten}, R.~I. {Epstein}, {Superfluid Dissipation Time Scales in Neutron
  Star Crusts}. \emph{Astrophys. J.} \textbf{342}, 951 (1989),
  \doi{10.1086/167652}.

\bibitem{Sedrakian1995}
A.~D. {Sedrakian}, D.~M. {Sedrakian}, {Superfluid Core Rotation in Pulsars. I.
  Vortex Cluster Dynamics}. \emph{Astrophys. J.} \textbf{447}, 305 (1995),
  \doi{10.1086/175876}.

\bibitem{Baym1969Natur}
G.~{Baym}, C.~{Pethick}, D.~{Pines}, M.~{Ruderman}, {Spin up in neutron stars:
  The future of the Vela pulsar}. \emph{Nature} \textbf{224}, 872--874 (1969),
  \doi{10.1038/224872a0}.

\bibitem{Cordes1988}
J.~M. {Cordes}, G.~S. {Downs}, J.~{Krause-Polstorff}, {JPL Pulsar Timing
  Observations. V. Macro- and Microjumps in the VELA Pulsar 0833-45}.
  \emph{Astrophys. J.} \textbf{330}, 847 (1988), \doi{10.1086/166518}.

\bibitem{BekarevichKhalatnikov1961}
I.~L. Bekarevich, I.~M. Khalatnikov, Phenomenological derivation of the
  equations of vortex motion in {He} {II}. \emph{Soviet Physics JETP}
  \textbf{13}, 643--646 (1961).

\bibitem{SedrakianSedrakian1995}
A.~D. Sedrakian, D.~M. Sedrakian, Superfluid hydrodynamics of rotating neutron
  stars. \emph{Soviet Physics JETP} \textbf{81}~(2), 382--389 (1995).

\bibitem{AlparSauls1988}
M.~A. Alpar, J.~A. Sauls, On the Dynamical Coupling Between the Superfluid
  Interior and the Crust of a Neutron Star. \emph{Astrophys. J.} \textbf{327},
  723--725 (1988), \doi{10.1086/166227}.

\bibitem{Anderson:1975zze}
P.~W. Anderson, N.~Itoh, {Pulsar glitches and restlessness as a hard
  superfluidity phenomenon}. \emph{Nature} \textbf{256}, 25--27 (1975),
  \doi{10.1038/256025a0}.

\bibitem{Link2014}
B.~{Link}, {Thermally Activated Post-Glitch Response of the Neutron Star Inner
  Crust and Core. I. Theory}. \emph{Astrophys. J.} \textbf{789}~(2), 141
  (2014), \doi{10.1088/0004-637X/789/2/141}.

\bibitem{Link:1999ca}
B.~Link, R.~I. Epstein, J.~M. Lattimer, {Pulsar constraints on neutron star
  structure and equation of state}. \emph{Phys. Rev. Lett.} \textbf{83},
  3362--3365 (1999), \doi{10.1103/PhysRevLett.83.3362}.

\bibitem{Chamel2013}
N.~{Chamel}, {Crustal Entrainment and Pulsar Glitches}. \emph{Phys. Rev. Lett.}
  \textbf{110}~(1), 011101 (2013), \doi{10.1103/PhysRevLett.110.011101}.

\bibitem{Almirante:2025cqe}
G.~Almirante, M.~Urban, {Superfluid Density in Linear Response Theory: Pulsar
  Glitches from the Inner Crust of Neutron Stars}. \emph{Phys. Rev. Lett.}
  \textbf{135}~(13), 132701 (2025), \doi{10.1103/mg61-gw93}.

\bibitem{Watanabe:2017nzj}
G.~Watanabe, C.~J. Pethick, {Superfluid Density of Neutrons in the Inner Crust
  of Neutron Stars: New Life for Pulsar Glitch Models}. \emph{Phys. Rev. Lett.}
  \textbf{119}~(6), 062701 (2017), \doi{10.1103/PhysRevLett.119.062701}.

\bibitem{Sedrakian:1998ki}
A.~Sedrakian, J.~M. Cordes, {Vortex-interface interactions and generation of
  glitches in pulsars}. \emph{MNRAS} \textbf{307}, 365–375 (1999),
  \doi{10.1046/j.1365-8711.1999.02638.x}.

\bibitem{Gugercinoglu:2014cda}
E.~G{\"u}gercino{\u{g}}lu, M.~A. Alpar, {Vortex Creep Against Toroidal Flux
  Lines, Crustal Entrainment, and Pulsar Glitches}. \emph{Astrophys. J. Lett.}
  \textbf{788}, L11 (2014), \doi{10.1088/2041-8205/788/1/L11}.

\bibitem{Hossain2022}
K.~Hossain, \emph{et~al.}, Rotating quantum turbulence in the unitary Fermi
  gas. \emph{Phys. Rev. A} \textbf{105}, 013304 (2022),
  \doi{10.1103/PhysRevA.105.013304},
  \url{https://link.aps.org/doi/10.1103/PhysRevA.105.013304}.

\bibitem{Poli2023}
E.~{Poli}, \emph{et~al.}, {Glitches in Rotating Supersolids}. \emph{Phys. Rev.
  Lett.} \textbf{131}~(22), 223401 (2023),
  \doi{10.1103/PhysRevLett.131.223401}.

\bibitem{Marmorini:2020zfp}
G.~Marmorini, S.~Yasui, M.~Nitta, {Pulsar glitches from quantum vortex
  networks}. \emph{Sci. Rep.} \textbf{14}~(1), 7857 (2024),
  \doi{10.1038/s41598-024-56383-w}.

\bibitem{Sedrakian:1998vi}
A.~Sedrakian, I.~Wasserman, J.~M. Cordes, {Precession of isolated neutron stars
  I: effects of imperfect pinning}. \emph{Astrophys. J.} \textbf{524}, 341
  (1998), \doi{10.1086/307777}.

\bibitem{Link2003}
B.~{Link}, {Constraining Hadronic Superfluidity with Neutron Star Precession}.
  \emph{Phys. Rev. Lett.} \textbf{91}~(10), 101101 (2003),
  \doi{10.1103/PhysRevLett.91.101101}.

\bibitem{Jones:2016oyh}
D.~I. Jones, G.~Ashton, R.~Prix, {Implications of the Occurrence of Glitches in
  Pulsar Free Precession Candidates}. \emph{Phys. Rev. Lett.}
  \textbf{118}~(26), 261101 (2017), \doi{10.1103/PhysRevLett.118.261101}.

\bibitem{Ruderman1970}
M.~{Ruderman}, {Long Period Oscillations in Rotating Neutron Stars}.
  \emph{Nature} \textbf{225}~(5233), 619--620 (1970), \doi{10.1038/225619a0}.

\bibitem{Noronha:2007qf}
J.~Noronha, A.~Sedrakian, {Tkachenko modes as sources of quasiperiodic pulsar
  spin variations}. \emph{Phys. Rev. D} \textbf{77}, 023008 (2008),
  \doi{10.1103/PhysRevD.77.023008}.

\bibitem{ChandlerBaym1986}
E.~Chandler, G.~Baym, Long-wavelength collective modes of a vortex lattice in a
  rotating superfluid. \emph{J. Low Temp. Phys.} \textbf{62}~(1--2), 119--142
  (1986), \doi{10.1007/BF00683444}.

\bibitem{CarterLanglois1995}
B.~Carter, D.~Langlois, Relativistic models for interacting superfluids.
  \emph{Nucl. Phys. B} \textbf{454}, 402--424 (1995),
  \doi{10.1016/0550-3213(95)00449-3}.

\bibitem{Langlois1998}
D.~Langlois, D.~M. Sedrakian, B.~Carter, Differential rotation of relativistic
  superfluid in neutron stars. \emph{MNRAS} \textbf{297}, 1189--1201 (1998),
  \doi{10.1046/j.1365-8711.1998.01574.x}.

\bibitem{Gusakov2016}
M.~E. Gusakov, Relativistic Formulation of the
  {{Hall-Vinen-Bekarevich-Khalatnikov}} Superfluid Hydrodynamics. \emph{Phys.
  Rev. D} \textbf{93}~(6), 064033 (2016).

\bibitem{Gusakov2016a}
M.~E. Gusakov, V.~A. Dommes, Relativistic Dynamics of
  Superfluid-Superconducting Mixtures in the Presence of Topological Defects
  and an Electromagnetic Field with Application to Neutron Stars. \emph{Phys.
  Rev. D} \textbf{94}~(8), 083006 (2016).

\bibitem{Rau2020}
P.~B. Rau, I.~Wasserman, Relativistic finite temperature multifluid
  hydrodynamics in a neutron star from a variational principle. \emph{Phys.
  Rev. D} \textbf{102}, 063011 (2020), \doi{10.1103/PhysRevD.102.063011}.

\bibitem{Alford2008}
M.~G. Alford, A.~Schmitt, K.~Rajagopal, T.~Sch{\"a}fer, Color superconductivity
  in dense quark matter. \emph{Rev. Mod. Phys.} \textbf{80}~(4), 1455--1515
  (2008), \doi{10.1103/RevModPhys.80.1455}.

\bibitem{Bailin1984}
D.~Bailin, A.~Love, Superfluidity and superconductivity in relativistic fermion
  systems. \emph{Phys. Rep.} \textbf{107}~(6), 325--385 (1984),
  \doi{10.1016/0370-1573(84)90145-5}.

\bibitem{Alford1999}
M.~G. Alford, K.~Rajagopal, F.~Wilczek, Color-flavor locking and chiral
  symmetry breaking in high density QCD. \emph{Nucl. Phys. B}
  \textbf{537}~(1-3), 443--458 (1999), \doi{10.1016/S0550-3213(98)00668-3}.

\bibitem{Kaplan2002}
D.~B. Kaplan, S.~Reddy, Novel phases and transitions in color flavor locked
  matter. \emph{Phys. Rev. D} \textbf{65}, 054042 (2002),
  \doi{10.1103/PhysRevD.65.054042}.

\bibitem{Bedaque2002}
P.~F. Bedaque, T.~Sch{\"a}fer, High density quark matter under stress.
  \emph{Nucl. Phys. A} \textbf{697}, 802--822 (2002),
  \doi{10.1016/S0375-9474(01)01219-3}.

\bibitem{Anglani2014}
R.~Anglani, \emph{et~al.}, Crystalline color superconductors. \emph{Rev. Mod.
  Phys.} \textbf{86}~(2), 509--561 (2014), \doi{10.1103/RevModPhys.86.509}.

\bibitem{Sedrakian:2003tr}
  A.~Sedrakian, ``Superconductivity with deformed Fermi surfaces and compact
stars,'' in \emph{Superdense QCD Matter and Compact Stars},
D.~Blaschke and D.~Sedrakian, Eds., vol.~197 of
\emph{NATO Advanced Research Workshop on Superdense QCD Matter and Compact Stars}
(Springer, 2003), pp.~209--224, \doi{10.1007/1-4020-3430-X_12}.

\bibitem{Grigorian2005}
H.~Grigorian, D.~Blaschke, D.~N. Voskresensky, Cooling of neutron stars with
  color superconducting quark cores. \emph{Phys. Rev. C} \textbf{71}, 045801
  (2005), \doi{10.1103/PhysRevC.71.045801}.

\bibitem{Warringa2012}
H.~J. Warringa, The phase diagram of neutral quark matter with pseudoscalar
  condensates in a magnetic field. \emph{Phys. Rev. D} \textbf{86}, 085029
  (2012), \doi{10.1103/PhysRevD.86.085029}.

\bibitem{Noronha2007}
J.~L. Noronha, I.~A. Shovkovy, Color-flavor locked superconductor in a magnetic
  field. \emph{Phys. Rev. D} \textbf{76}, 105030 (2007),
  \doi{10.1103/PhysRevD.76.105030}.

\bibitem{Alford2010}
M.~G. Alford, A.~Sedrakian, Color-magnetic flux tubes in quark matter cores of
  neutron stars. \emph{J. Phys. G: Nucl. Part. Phys.} \textbf{37}, 075202
  (2010), \doi{10.1088/0954-3899/37/7/075202}.

\bibitem{Balachandran:2005ev}
A.~P. Balachandran, S.~Digal, T.~Matsuura, {Semi-superfluid strings in high
  density QCD}. \emph{Phys. Rev. D} \textbf{73}, 074009 (2006),
  \doi{10.1103/PhysRevD.73.074009}.

\bibitem{STAR:2017ckg}
{STAR Collaboration}, Global $\Lambda$ hyperon polarization in nuclear
  collisions. \emph{Nature} \textbf{548}, 62--65 (2017),
  \doi{10.1038/nature23004}.

\bibitem{Cipriani:2012hr}
M.~Cipriani, W.~Vinci, M.~Nitta, {Colorful boojums at the interface of a color
  superconductor}. \emph{Phys. Rev. D} \textbf{86}, 121704 (2012),
  \doi{10.1103/PhysRevD.86.121704}.

\bibitem{Chatterjee2018}
C.~Chatterjee, M.~Nitta, S.~Yasui, Quark-hadron continuity under rotation:
  Vortex continuity or boojum? \emph{Phys. Rev. D} \textbf{99}, 034001 (2019),
  \doi{10.1103/PhysRevD.99.034001}.

\bibitem{Alford2019}
M.~G. Alford, G.~Baym, K.~Fukushima, T.~Hatsuda, M.~Tachibana, Continuity of
  vortices from the hadronic to the color-flavor locked phase in dense matter.
  \emph{Phys. Rev. D} \textbf{99}~(3), 036004 (2019),
  \doi{10.1103/PhysRevD.99.036004}.

\bibitem{Eto2014}
M.~Eto, Y.~Hirono, M.~Nitta, S.~Yasui, Vortices and other topological solitons
  in dense quark matter. \emph{Prog. Theor. Exp. Phys.} \textbf{2014}, 012D01
  (2014), \doi{10.1093/ptep/ptt095}.

\bibitem{Mannarelli2007}
M.~Mannarelli, K.~Rajagopal, R.~Sharma, The rigidity of crystalline color
  superconducting quark matter. \emph{Phys. Rev. D} \textbf{76}, 074026 (2007),
  \doi{10.1103/PhysRevD.76.074026}.

\end{thebibliography}


\section*{Acknowledgments}

\paragraph*{Funding:}
 A.\,S.   was funded by the Deutsche For\-schungs\-gemeinschaft Grant
 No. SE 1836/6-1 and the Polish NCN Grant No. 2023/51/B/ST9/02798.\\
 P.\,B.\,R. is supported by the Simons Foundation through a SCEECS
 postdoctoral fellowship (grant No. PG013106-02).
 
 \paragraph*{Author contributions:}
Armen Sedrakian: Writing - original draft, Writing - review \& editing,
Peter Rau:  Writing - original draft,  Writing - review \& editing

\paragraph*{Competing interests:}
There are no competing interests to declare.

\paragraph*{Data, code, and materials availability:}
All data and code needed to evaluate and reproduce the results in the paper are present in the paper.
Data sets generated during the current study are available from the
corresponding author on reasonable request. 

\end{document}